\documentclass{article}
\usepackage[preprint]{colm2026_conference}
\usepackage{microtype,hyperref,url,booktabs,array,amsmath,amssymb,xcolor,graphicx,multirow,placeins,capt-of,float}
\usepackage[T1,OT1]{fontenc}
\usepackage{listings,needspace}
\definecolor{inkblue}{RGB}{30,65,100}
\hypersetup{colorlinks=true,linkcolor=inkblue,citecolor=inkblue,urlcolor=inkblue,
  pdftitle={Qwen-Audio-3.1-Realtime: Towards Reliable Agentic Voice Interaction},
  pdfauthor={Lujia Bao, Qian Chen, Luyao Cheng, Chong Deng, Yuxiang Kong, Xiangang Li, Xu Li, Jiaqing Liu, Chao-Hong Tan, Haoyu Wang, Wen Wang, Xilou Wang, Haoxiang Xu, Junhao Xu, Liang Yi, Binbin Zhang, Qinglin Zhang, Qiquan Zhang},
  pdfsubject={Technical report},
  pdfkeywords={Qwen-Audio, spoken dialogue, voice agents, audio reasoning, tool use}}
\newcommand{\model}{Qwen-Audio-3.1-Realtime}
\newcommand{\mopd}{M\textsuperscript{2}-OPD}
\newcommand{\metricup}[1]{\mbox{#1}}
\newcommand{\metricdown}[1]{\mbox{#1\ensuremath{\downarrow}}}
\newcommand{\tablefont}{\scriptsize}
\newcommand{\theadcenter}[1]{\raisebox{\dimexpr-0.5\height+0.5\depth\relax}{\shortstack[c]{#1}}}
\title{Qwen-Audio-3.1-Realtime: Towards Reliable Agentic Voice Interaction}
\author{\makebox[\dimexpr\textwidth-2\tabcolsep\relax][c]{\textnormal{Alibaba Token Foundry, Alibaba Group}}}
\begin{document}
\maketitle
\lhead{Qwen-Audio-3.1-Realtime Technical Report}
\begin{abstract}
Real-time voice assistants must reason over evolving requests, execute actions, and follow conversational rules. \model{} brings these requirements together through \emph{Think}, \emph{Act}, and \emph{Speak and Coordinate}. Think combines Core-Cocktail supervised fine-tuning with Multimodality and Multi-Teacher On-Policy Distillation (\mopd{}) to transfer language capabilities and develop native audio skills. Act uses self-evolving executable environments and multi-granularity rollouts for Group Relative Policy Optimization (GRPO), teaching the model to use tools, interpret feedback, and complete tasks. Speak and Coordinate aligns whether, when, and how the assistant speaks or acts. We evaluate audio reasoning, multilingual understanding, tool use, conversational behavior, full-duplex interaction, and safety. Compared with Qwen-Audio-3.0-Realtime, 3.1 raises overall task success from 78.4\% to 82.0\% on our half-duplex speech-to-text adaptation of $\tau$-Voice. On speech-to-speech Full-Duplex-Bench v1.5, the response rate to background speech falls from 73.0\% to 13.0\%. We also present a separate Voice Harness prototype, using Qwen-Audio-3.0-Realtime as its foreground, that extends spoken interaction to persistent tasks through foreground--background coordination and memory.
\end{abstract}

\begin{center}
\begin{minipage}{\textwidth}
  \centering
  \includegraphics[width=\textwidth]{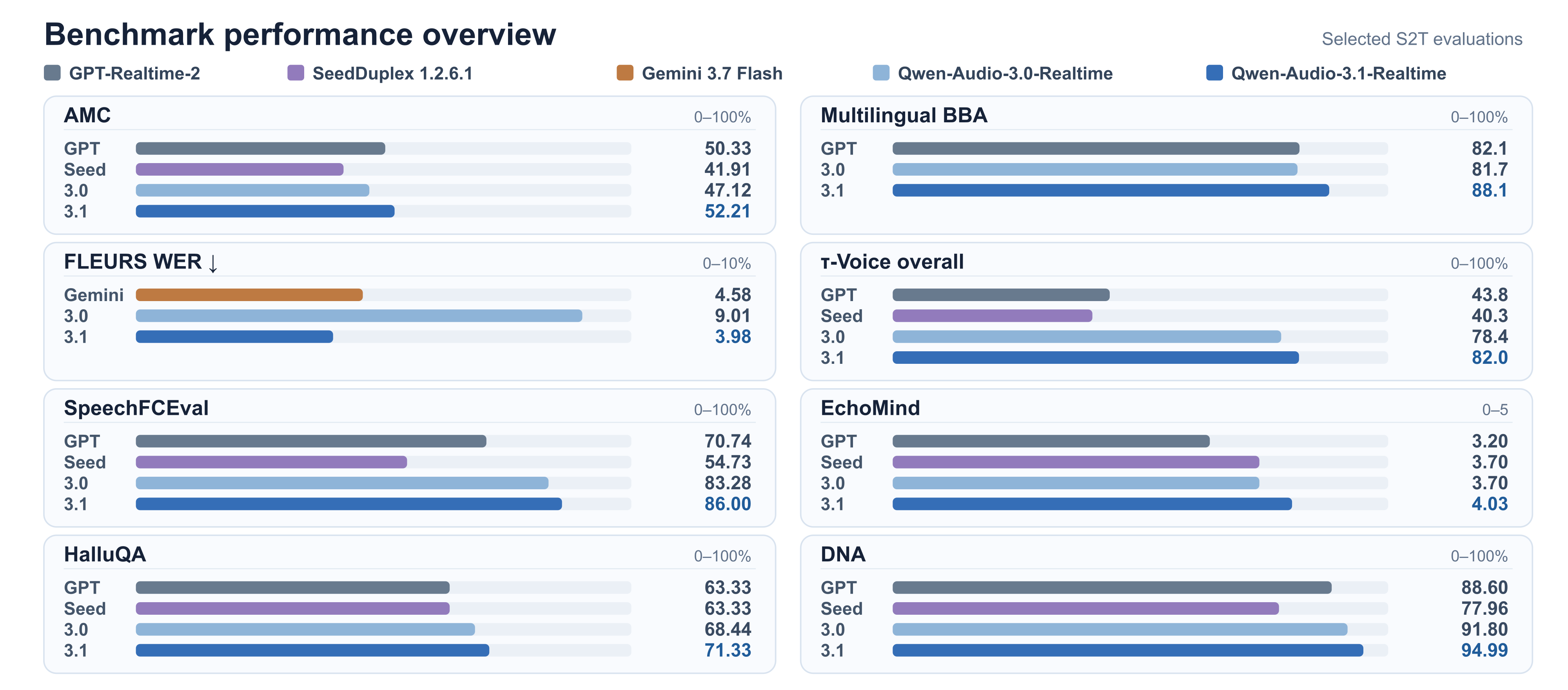}
  \captionof{figure}{Selected S2T results, showing only models with a reported score in each panel. Panels use independent printed ranges: FLEURS WER is lower-is-better, EchoMind is scored 1--5, and other results are percentages. See Section~\ref{sec:evaluation} for benchmark-specific protocols and citations.}
  \label{fig:benchmark-overview}
\end{minipage}
\end{center}

\section{Introduction: Realtime Beyond Latency}
Qwen-Audio-3.0-Realtime \citep{qwenaudio30realtime} provides the starting point: real-time listening and speaking. Version 3.1 investigates what is needed when spoken interaction becomes a setting for reasoning and action. An assistant must retain constraints as requests evolve, choose and execute tools, interpret feedback, and follow the user's instructions and interaction rules. Low response latency is one requirement within this broader problem.

Consider a user who requests a route, adds constraints, asks the assistant to wait until they finish, and later changes the destination while a query is running. Appropriate behavior depends on understanding the goal, tracking execution, deciding whether to speak, and communicating only what the available evidence supports. This scenario motivates the interaction design developed in this report.

We organize \model{} around three layers. \textbf{Think} provides the foundation for audio reasoning and multi-turn instruction following through supervision from the source Text LLM and audio-conditioned signals. \textbf{Act} converts spoken intent into executable decisions under task and policy constraints. \textbf{Speak and Coordinate} governs whether, when, and how the assistant speaks or acts. These layers support concurrent action and speech, such as reporting progress while a tool runs.

Action learning is the methodological center of this report. We describe executable domains containing tools, policies, and databases, together with associated tasks; construction checks and rollout-driven evolution; and rollout feedback at multiple granularities for Group Relative Policy Optimization (GRPO) \citep{deepseekmath}. This treatment extends beyond function-call syntax to state verification, long-horizon task completion, appropriate refusal, and progress communication.

Following the architecture--training--evaluation structure of \citet{funaudiochat}, the report makes three main contributions. First, it presents a foundation post-training pipeline that combines Core-Cocktail supervised fine-tuning (SFT) with the Multimodality OPD and Multi-Teacher OPD paths, connecting text-grounded reasoning to native audio capabilities. Second, it develops executable, self-evolving training environments with rollout-based feedback at dialogue, milestone, and turn levels for learning tool use, task completion, and grounded progress communication. Third, it formulates spoken interaction as a conversational policy over whether, when, and how to speak or act. The report also presents a separate Voice Harness design with explicit task and memory boundaries. Evaluation follows the same intelligence, action, interaction, and real-time organization.

\section{Qwen-Audio-3.1-Realtime Overview}
\begin{center}
\begin{minipage}{\textwidth}
  \centering
  \includegraphics[width=\textwidth]{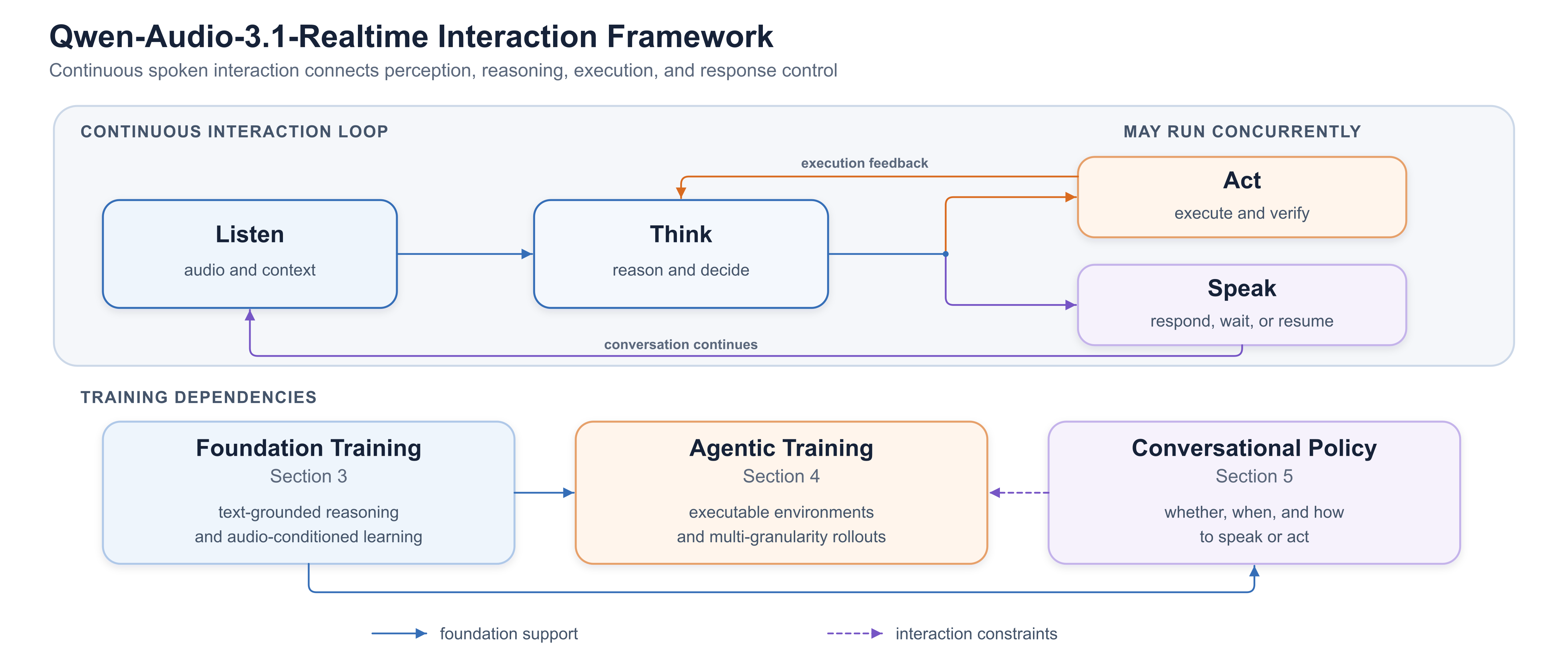}
  \captionof{figure}{Runtime interaction and training dependencies. Foundation training (Section~\ref{sec:foundation}) supports agentic training (Section~\ref{sec:act}) and conversational policy alignment (Section~\ref{sec:conversational-policy}). Solid arrows between training boxes denote foundation support; the dashed arrow denotes interaction constraints. The upper loop branches from Think to Act and Speak, which can run concurrently.}
  \label{fig:interaction-loop}
\end{minipage}\end{center}
Foundation training first establishes an audio policy. Agentic learning and conversational alignment build on this policy and share interaction constraints, including appropriate acknowledgments and truthful progress reporting. Figure~\ref{fig:interaction-loop} separates these dependencies from runtime behavior: Listen and Think support concurrent Act and Speak, allowing progress updates while tools execute.

\subsection{Architecture and interfaces}
Figure~\ref{fig:model-framework} shows how response generation, interaction control, and voice rendering work together during continuous spoken dialogue. Both the full-duplex decision model and the speech-to-text model use an audio encoder followed by a large language model (LLM). The former predicts whether the system should continue listening, begin speaking, stop, or resume; the latter generates the response content as text. The context-aware voice renderer combines this content with conversation history, voice cues, and acoustic context to produce streaming speech, while full-duplex decisions govern when speech begins, stops, or resumes. The two Audio Encoder + LLM paths share the same high-level architecture for interaction decisions and response generation.

\begin{center}
\begin{minipage}{\textwidth}
  \centering
  \includegraphics[width=\textwidth]{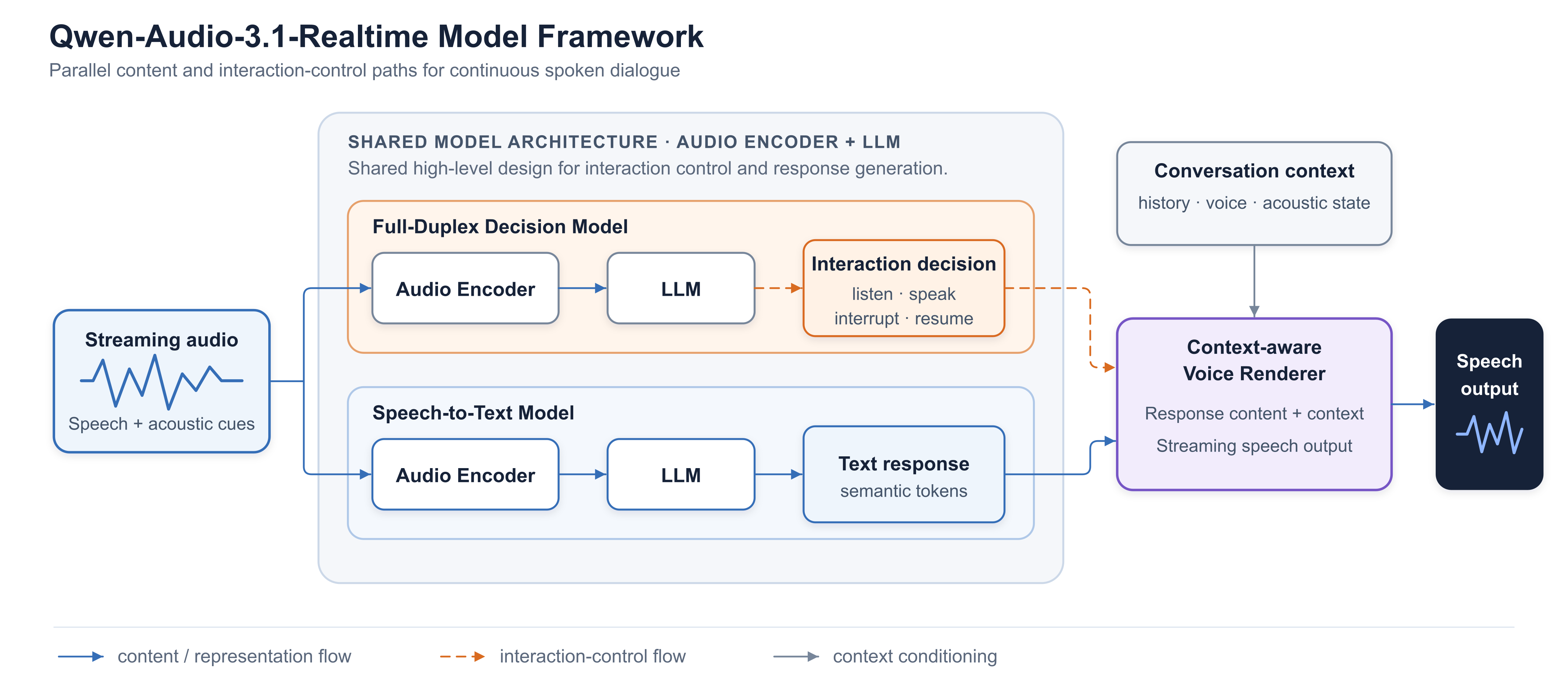}
  \captionof{figure}{Qwen-Audio-3.1-Realtime model framework for continuous spoken interaction. Streaming audio is processed by two models with the same Audio Encoder + LLM architecture. The full-duplex path controls speech timing, while the Speech-to-Text path generates response content. The context-aware voice renderer produces streaming speech from this content and the conversation and acoustic context, guided by the interaction decisions.}
  \label{fig:model-framework}
\end{minipage}\end{center}

\subsection{Release scope}
We compare the released Qwen-Audio-3.0-Realtime and Qwen-Audio-3.1-Realtime systems. Results are interpreted within the modality, benchmark, and judge settings stated in Section~\ref{sec:evaluation} and the table captions.

\section{Foundation Training for Spoken Intelligence}
\label{sec:foundation}
The foundation recipe combines text-grounded and audio-conditioned supervision for an audio language model (AudioLM). It builds on the Core-Cocktail post-training strategy introduced in DrVoice and further developed in Fun-Audio-Chat \citep{drvoice,funaudiochat}. Training has four stages: Core-Cocktail produces an initial audio policy; Multimodality OPD supervises its audio-conditioned trajectories with distributions from a Text Teacher and Audio Reference; domain GRPO produces specialists; and Multi-Teacher OPD consolidates their supervision into the main model.

\begin{figure*}[t]
  \centering
  \includegraphics[width=\textwidth]{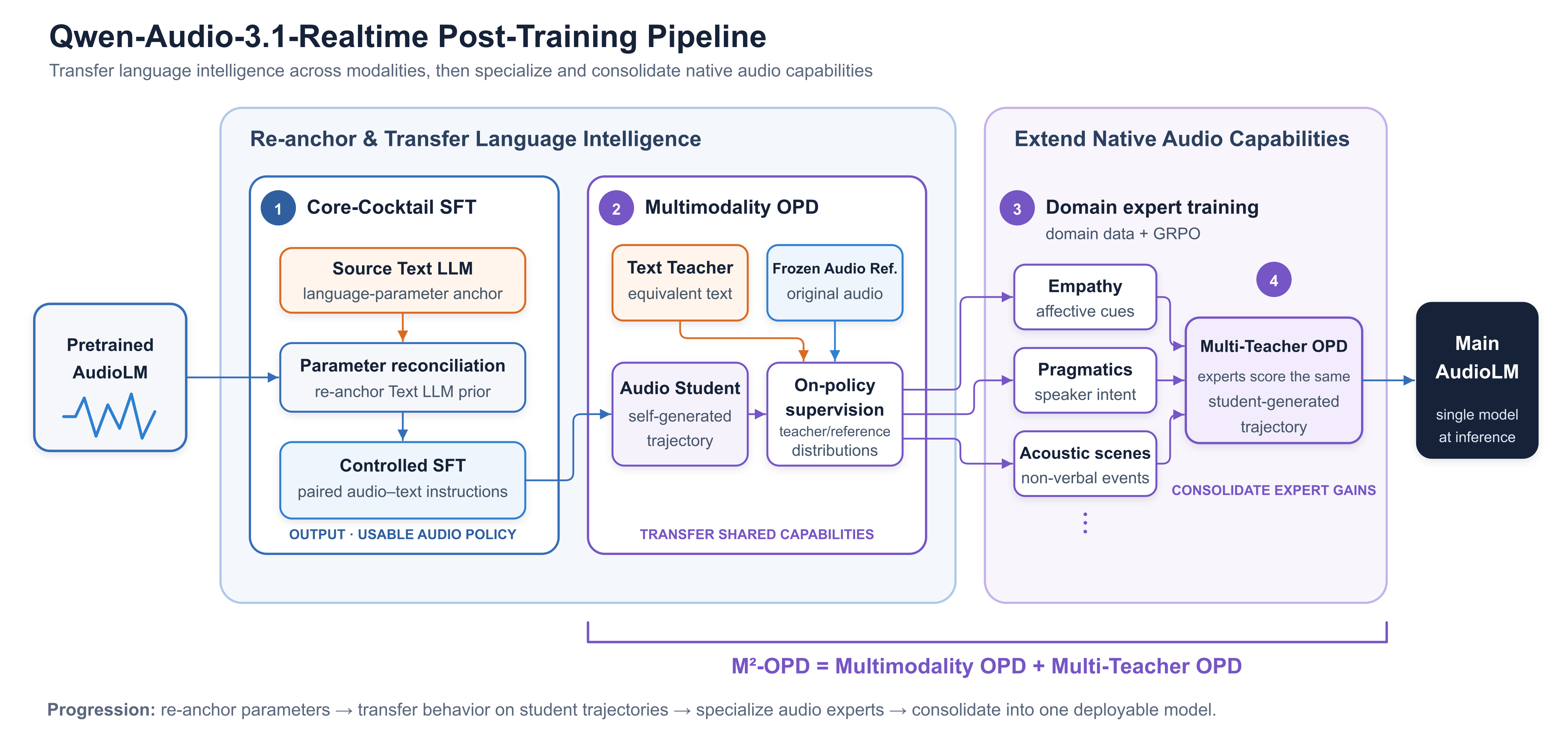}
  \caption{Qwen-Audio-3.1-Realtime post-training pipeline. Core-Cocktail SFT combines parameter reconciliation with controlled audio-text instruction tuning. Multimodality OPD transfers cross-modal capabilities from a Text Teacher while a frozen Audio Reference supplies audio-conditioned supervision. Domain experts specialize with domain data and GRPO; the ellipsis denotes additional domains. Multi-Teacher OPD consolidates their token-level supervision on student-generated trajectories into a single deployable model. The two OPD paths together form \mopd{}\@.}
  \label{fig:post-training-pipeline}
\end{figure*}

We use \mopd{} to refer to the two on-policy distillation (OPD) stages together. Both operate on student-generated trajectories \citep{lu2025opd}. Figure~\ref{fig:post-training-pipeline} shows how text-grounded supervision and native-audio specialization enter the same post-training pipeline. Only the consolidated main model is required at inference time.

\subsection{Core-Cocktail supervised fine-tuning}
The audio-pretrained AudioLM follows the three-stage recipe in the Qwen-Audio-3.0-ASR Technical Report \citep[Sections~3.1--3.3]{bian2026qwenaudioasr}. The encoder first learns speech representations through self-supervised pretraining and supervised encoder--decoder training on paired audio-text data. After discarding the temporary decoder, we connect the frozen encoder to a pretrained Qwen LLM through an adapter and train the adapter and LLM for audio-text alignment. Finally, cooldown adaptation uses higher-quality, rebalanced multilingual and multi-task data to refine alignment and instruction following.

Core-Cocktail is the cold-start stage of post-training. It takes this audio-pretrained AudioLM and uses the source Text LLM as a language-parameter anchor. Building on the Core-Cocktail strategy \citep{drvoice,funaudiochat}, the pipeline reconciles their shared language parameters before controlled SFT establishes a usable audio policy. Within this stage, audio and text instruction data jointly supervise audio behavior and language-centered tasks.

\paragraph{Training scope.}
Following the broad multi-domain post-training formulation of Fun-Audio-Chat \citep{funaudiochat}, the foundation recipe uses million-hour-scale paired audio-text data. The data span general audio understanding, spoken instruction following, multi-turn interaction, tool-oriented requests, emotion and empathy, pragmatic intent, and acoustic-scene understanding.

\subsection{Multimodality OPD with an Audio Reference}
Let $\pi_\theta$ denote the Audio Student policy, $x_a$ an audio input, and $x_t$ its semantically equivalent text on tasks where such a representation is sufficient. The student samples a trajectory $y=(y_1,\ldots,y_J)$ as
\begin{equation}
y\sim\pi_\theta(\,\cdot\mid x_a).
\end{equation}
Here, $J$ is the sampled response length and $y_{<j}=(y_1,\ldots,y_{j-1})$. At token position $j\in\{1,\ldots,J\}$, the Audio Student distribution on this prefix is $p_\theta^{(j)}=\pi_\theta(\,\cdot\mid x_a,y_{<j})$. The Text Teacher receives the equivalent text prompt and the same student prefix, providing
\begin{equation}
q_T^{(j)}=q_T(\,\cdot\mid x_t,y_{<j}).
\end{equation}
A frozen Audio Reference, initialized from the Audio Student at the start of this stage, receives the original audio context and the same prefix,
\begin{equation}
q_{\mathrm{ref}}^{(j)}=q_{\mathrm{ref}}(\,\cdot\mid x_a,y_{<j}).
\end{equation}
The student, teacher, and reference distributions are therefore evaluated on the same Audio Student trajectory. The Text Teacher does not provide a pre-generated answer; it scores the student-generated prefix at each position. OPD updates $p_\theta^{(j)}$ using two complementary supervisory distributions: $q_T^{(j)}$ provides supervision for knowledge, reasoning, instructions, and tool decisions, while $q_{\mathrm{ref}}^{(j)}$ supplies an audio-conditioned reference. The alignment target is behavior along the generated trajectory rather than input-feature similarity alone.

This path applies when relevant task information can be represented in text. For tasks shaped by emotion, pragmatic intent, or acoustic context, audio-conditioned supervision supplies a learning signal for information beyond the transcript.

\subsection{Multi-Teacher OPD}
The pipeline first uses domain data and GRPO to produce specialists across domains, including empathy, pragmatic intent, and acoustic scenes. These are examples rather than an exhaustive list, as indicated by the ellipsis in Figure~\ref{fig:post-training-pipeline}. This specialist-training stage forms domain-specific teachers; it does not itself merge their behaviors into the main model. Multi-Teacher OPD is the subsequent consolidation stage. The main AudioLM generates its own audio-conditioned trajectories; for every student prefix, each relevant expert receives the same audio context and supplies a token-level distribution. If $k$ indexes the audio-domain experts, the distribution from expert $k$ at position $j$ is
\begin{equation}
q_k^{(j)}=q_k(\,\cdot\mid x_a,y_{<j}).
\end{equation}
The student distribution remains $p_\theta^{(j)}=\pi_\theta(\,\cdot\mid x_a,y_{<j})$, now supervised by the distributions of the experts relevant to that domain. Unlike Multimodality OPD, which pairs a Text Teacher with an Audio Reference, this stage draws supervision from audio-domain experts on the same student trajectory.

The domain teachers participate only during post-training, and consolidation produces a single main AudioLM policy for deployment. The domain-specific GRPO in this section trains audio specialists for Multi-Teacher OPD. Section~\ref{sec:act} describes a separate use of GRPO with executable tasks to train the main policy's action behavior.

\begin{table*}[t]
\centering\tablefont
\begin{tabular}{>{\raggedright\arraybackslash}m{0.16\textwidth}>{\raggedright\arraybackslash}m{0.24\textwidth}>{\raggedright\arraybackslash}m{0.32\textwidth}>{\raggedright\arraybackslash}m{0.14\textwidth}}
\toprule
Stage & Objective & Data and supervision & Stage output \\
\midrule
Core-Cocktail SFT & Re-anchor the AudioLM to the source Text LLM and establish a usable audio instruction policy. & Million-hour-scale multi-domain paired audio-text data; controlled SFT after parameter reconciliation. & Initial audio policy. \\
\addlinespace[2pt]
Multimodality OPD & Transfer text-grounded reasoning, instruction following, and tool decisions while retaining audio-conditioned behavior. & Audio Student trajectories receive token-level distributions from a Text Teacher and a frozen Audio Reference. & Audio policy after cross-modal transfer. \\
\addlinespace[2pt]
Domain expert training & Learn native-audio behaviors that depend on acoustic and paralinguistic signals. & Domain-specific audio data and rewards; GRPO for domains such as empathy, pragmatic intent, and acoustic scenes. & Audio-domain specialists. \\
\addlinespace[2pt]
Multi-Teacher OPD & Consolidate native-audio specializations into one deployable AudioLM\@. & Audio-domain specialists supervise the main model's student-generated trajectories. & Unified main AudioLM\@. \\
\bottomrule
\end{tabular}
\caption{Foundation post-training stages, objectives, supervision interfaces, and stage outputs. Domain expert training creates the teachers consumed by Multi-Teacher OPD; the latter performs consolidation into the main model.}
\label{tab:foundation-stages}
\end{table*}

\section{Learning to Act in Executable Environments}
\label{sec:act}
\begin{figure*}[t]
  \centering
  \includegraphics[width=\textwidth]{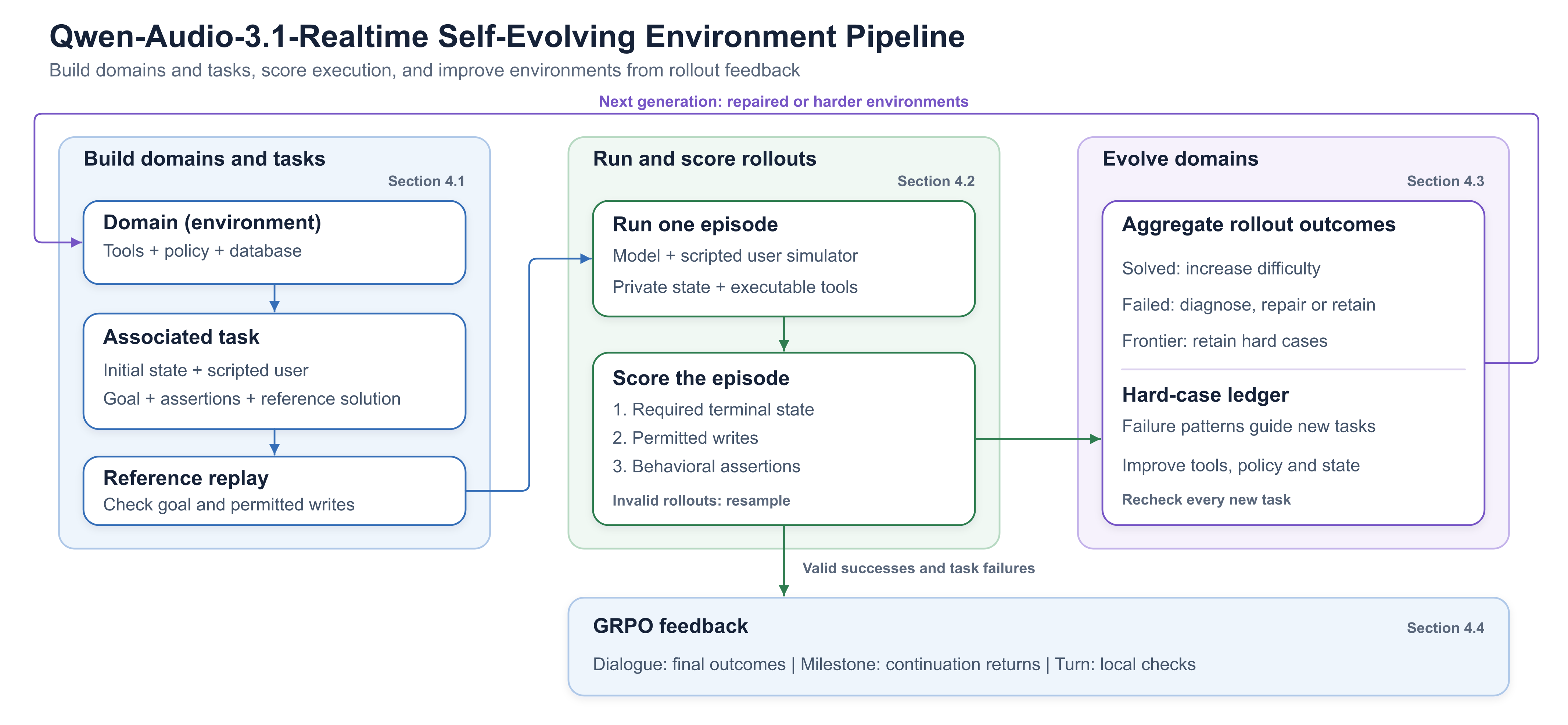}
  \caption{Executable domain construction, rollout scoring, and self-evolution. Each domain contains tools, a database, and policies; associated tasks specify individual episodes. Reference replay checks task consistency before rollout. Completed episodes are scored for terminal state, permitted writes, and behavioral requirements. Aggregated outcomes guide repairs, harder environments, and frontier tasks; valid successes and failures supply GRPO signals at three granularities.}
  \label{fig:self-evolving-pipeline}
\end{figure*}

The goal is to teach the model to complete tasks reliably under changing spoken requests and task rules. Producing a valid function call is only one step. The model may also need to identify a spoken parameter, check a record, ask for confirmation, update a database, and report what happened. We describe four steps: define executable domains and tasks (Section~\ref{sec:act:build}); run and score rollouts (Section~\ref{sec:act:reward}); use rollout feedback to improve domain quality and difficulty (Section~\ref{sec:act:evolve}); and provide dialogue-, milestone-, and turn-level reward signals for GRPO (Section~\ref{sec:act:gran}). Figure~\ref{fig:self-evolving-pipeline} summarizes their connections.

\subsection{Executable domains and tasks}
\label{sec:act:build}
An executable \emph{domain}, or environment, contains a tool pool, a stateful database, and business policies. A \emph{task} specifies an episode within that domain: an initial state, a user script, an expected terminal condition, behavioral assertions, and a reference tool-call chain. For example, a flight-service domain may support search and fare holds; one task asks the model to place a hold after confirming its details. Appendix~\ref{app:env-anatomy} develops this example. Each rollout uses a fixed version of the domain. Evolution changes the domain and its associated tasks between generations (Section~\ref{sec:act:evolve}).

\paragraph{Inside an environment.}
The \emph{business policy} is a natural-language document that enters the model's system prompt and states both what the desk can do and what it must never do; negative constraints (for example, ``this desk never takes payment details'') are first-class test targets. \emph{Tools} are plain functions over the episode's state: their JSON schemas are derived mechanically from their signatures, and failures are returned as tool results so the model can revise its next action. The \emph{database} is a JSON document whose \emph{schema} is shared by all tasks in the environment, while state is never shared across episodes---every episode starts from a fresh copy of the task's initial state, and state checks use that episode's private copy. Nor does every task write to the database: informational and refusal tasks leave the state untouched by design, and ``no change'' is a first-class expected outcome. 

Environments are seeded from a large corpus of open-source tool and Model Context Protocol (MCP) server definitions---ranging from individual tool schemas to full server tool sets---which supplies diverse, realistic domain topics and tool pools at scale. A code agent expands each selected seed into tools, a policy, and a database, and generates associated tasks under a fixed set of construction rules. Appendix~\ref{app:env-anatomy} illustrates these components and one associated task.

\paragraph{Inside a task.}
Each task instantiates one episode from the environment. The \emph{user script} specifies the user's identity, goal, the facts available to the user, and when to reveal them. The \emph{initial state} fixes the database copy the episode runs on. \emph{Behavioral assertions} name requirements that state cannot express, such as confirming details before a write. For state-changing and boundary-testing tasks, the \emph{expected terminal condition} distinguishes three outcomes: a \emph{write}, where the request is policy-compliant and the episode must reach a required state change; a \emph{justified refusal}, where the capability exists but policy forbids the request or its preconditions are unmet, so the correct behavior is to refuse with a reason and leave the state unchanged; or \emph{unsupported}, where the capability does not exist in the tool pool at all, so the correct behavior is to state the boundary truthfully. Asking for clarification is a move \emph{inside} an episode, not a terminal outcome; purely informational requests are scored by behavioral assertions rather than state change. 

\paragraph{Construction checks.}
Before rollout, construction checks replay the \emph{reference solution} from the initial state without a model. This verifies that the tools can reach the expected terminal condition and establishes the permitted write set. Additional checks cover the policy clauses under test and the separation of user-visible information from internal state. Reference replay checks executable feasibility; dialogue requirements, such as obtaining confirmation, are assessed from the model's rollout (Section~\ref{sec:act:reward}). Inconsistent tasks are repaired or discarded before training.

\subsection{Running and scoring a rollout}
\label{sec:act:reward}

\paragraph{Running one episode.}
Each rollout pairs the model with an LLM-based \emph{user simulator} that follows the task's script. The simulator sees the dialogue, but not the model's tool calls or internal state. In transactional or outbound scenarios, it may have user-side tools, such as checking a balance or making a payment; these tools update the same episode database.

The environment executes model tool calls against the private state and returns results or errors. The simulator ends the exchange as specified by its script, for example by hanging up in a phone-service task. Turn and tool-call limits also bound execution, and milestone checks can stop a rollout when a required intermediate state becomes unreachable.

\paragraph{Scoring after the episode.}
Scoring follows three checks in order. \emph{Terminal-state checks} compare the episode's private final state with the task's expected condition. \emph{Write checks} reject changes outside the permitted write set. \emph{Behavioral assertions} then check the dialogue and tool trace for requirements such as verifying information, obtaining confirmation before a write, and reporting status truthfully. Assertions record violations only and are skipped when their preconditions do not apply.

The evaluator uses recorded tool errors and state changes rather than inferring execution from the dialogue. A verdict requires a complete tool trace and all required judgments. A fluent reply cannot compensate for a failed state check or an unauthorized write; conversely, reaching the correct state does not excuse a violated interaction rule.

\subsection{Evolving domains from rollout results}
\label{sec:act:evolve}
For each task, the pipeline aggregates verdicts and traces from several scored rollouts and labels it consistently solved, consistently failed, or near the model's current capability boundary. Rollouts that cannot be scored at all---infrastructure failures, malformed traces, missing judgments---are \emph{invalid}: they are resampled rather than assigned an artificial zero reward, and do not affect the label. 

The task label and recorded failures guide the next generation. If the model repeatedly solves a task, the pipeline adds policy dependencies or state constraints to increase difficulty. If the model repeatedly fails, the pipeline first checks the environment itself: broken environments are repaired, while valid tasks that expose a model weakness are kept as training data. Tasks near the capability boundary are retained as frontier cases. A hard-case ledger records their failure patterns and guides later task generation. Before training, every new task undergoes construction checks, rollout checks, and difficulty assessment.

Feedback operates at two levels. At the task level, useful difficulty patterns are reused to create later tasks, while coverage targets track complexity and interaction length. At the framework level, repeated failures in task construction checks suggest changes to environment templates and the check rules or code. The pipeline therefore improves both the training tasks and the system that produces them.

\subsection{Three levels of reward signals for GRPO}
\label{sec:act:gran}
Group Relative Policy Optimization (GRPO) receives feedback at dialogue, milestone, and turn levels. Each comparison uses independent copies of the relevant task state and the same executable environment---tools, schema, and policies. State and trace checks apply where relevant, but the scored unit differs: a complete dialogue, a continuation from a saved milestone, or a local turn decision.

\paragraph{Dialogue level.}
The pipeline samples several complete conversations for the same task and compares their final outcomes. This directly trains task completion and policy compliance. It also provides the pass rates used for difficulty assessment (Section~\ref{sec:act:evolve}). The limitation is that a final reward does not always reveal which intermediate decision caused the result.

\paragraph{Milestone level.}
A reference tool chain defines important intermediate states, or milestones. Each milestone requires the expected successful calls, in order, with the required arguments; a failed call does not count. When a rollout reaches one, the pipeline saves the dialogue and database state, then samples several continuations from that shared point. Their final returns therefore compare the next decisions more directly.

A partial trajectory may already satisfy some requirements, make some violations irreversible, leave other obligations pending, or render some conditions irrelevant. The scorer keeps these cases separate: it waits for a pending action's deadline and never rewards a claim of completion before it is true.

\paragraph{Turn level.}
At a selected decision point, the model proposes an utterance, tool calls, or both. Local checks score tool choice, parameter grounding, unauthorized writes, and reply quality. Saved prefixes and database states, including states from earlier failures, can be reused to create these examples. This gives focused correction without replaying a complete conversation. Distractor tools and additional policy context control task difficulty. Before training, we filter tool names, dialogue fragments, and policy text that overlap with evaluation data.

The three levels provide feedback at different scales: complete tasks, intermediate decisions, and individual turns. They draw on the same underlying executable checks where applicable and can be combined across training stages.

\subsection{Speech-conditioned tool use and retrieval}
The pipeline can synthesize a user turn into speech and send the audio to the model. This makes spoken numbers, similar-sounding names, and negation part of tool selection and parameter grounding. Synthetic speech matches the deployment modality, but robustness to real recordings must still be evaluated separately.

Search training separates whether to search, what to query, and how to answer from evidence. We mix search-needed and no-search requests. For search-needed cases, the data pipeline estimates an appropriate count between one and five, removes near-duplicates, and keeps complementary queries, adding the current date when freshness matters. The model then answers from retrieved evidence. For a question such as ``What is chlamydia?'', ``chlamydia meaning'' and ``chlamydia definition'' add little beyond the original wording.

The reward follows the same split. The tool-call turn first scores whether search was needed. For a search call, a judge scores the queries on accuracy, coverage, specificity, and diversity; their mean is $q$. Independently of the number of submitted queries, the judge also estimates a reasonable count $n_{\mathrm{ref}}\in\{1,\ldots,5\}$ for the request. If the model proposes $n_{\mathrm{pred}}\geq1$ queries, their quality-and-count reward is
\begin{equation}
r_{\mathrm{query}}=q\min\!\left(1,\frac{n_{\mathrm{ref}}}{n_{\mathrm{pred}}}\right).
\label{eq:query-count-reward}
\end{equation}
The reward equals $q$ at or below the estimated count and falls in proportion to $n_{\mathrm{ref}}/n_{\mathrm{pred}}$ above it. A missing search call is handled by the separate search-decision reward. The final turn scores faithfulness, usefulness, safety, and spoken expression. This count penalty complements data-side pruning; it does not directly measure answer quality or latency.

\subsection{Reporting progress before tool calls}
Before a tool call, the model may briefly tell the user that work is starting, but it must not claim success before the task completes. Each tool-call turn is labeled as \emph{required}, \emph{disallowed}, or \emph{optional}; higher-level rules, such as acknowledging only the first call, are converted into these turn-level labels.

The evaluator checks whether the acknowledgment occurs in the correct place and, if it does, its length, repetition, language consistency, and implementation-detail disclosure. It separately identifies the first acknowledgment and the first substantive answer. This connects action learning with conversational policy: the action layer decides what to do, while the interaction policy decides how to communicate progress. Premature promises, repeated filler, and unnecessary internal details are penalized.

\section{Conversational Policy Alignment}
\label{sec:conversational-policy}
Conversational policy organizes whether, when, and how the assistant should speak or act. Tool-training details remain in Section~\ref{sec:act}; this section explains the constraints they share with dialogue behavior.

\subsection{Whether to speak or act: boundaries and permissions}
The policy must choose among answering directly, calling a tool, asking for clarification or confirmation, refusing, and remaining silent. Tool triggers, uncertainty, verification, and capability boundaries determine which choice is appropriate. Capability-boundary evaluation checks whether each reply remains consistent with a fixed capability configuration; it does not establish that a described capability is available. Learned policy following must be distinguished from permissions enforced by tools and the runtime. A spoken promise is neither proof of execution nor an authorization token.

\subsection{When to speak: waiting, yielding, and resuming}
The duplex decision policy conditions a decision on observed audio, history, and system state:
\begin{equation}
a_t=\pi(o_{\leq t},h_{\leq t},s_t).
\end{equation}
Here, $o_{\leq t}$ is the streaming audio observed through time $t$, $h_{\leq t}$ is the dialogue and interaction-event history, and $s_t$ is the current system state, including whether speech or tool execution is active. The action space distinguishes continuing to listen, beginning a response, stopping playback, and resuming after an interruption. History can contain temporary agreements, such as waiting for a completion signal. For turn timing, an explicit current-session agreement takes priority over scene-level defaults; tool permissions and safety rules remain binding. Backchannels, speech addressed to another participant, and genuine interruptions require different responses. Stopping playback and deciding whether to answer are distinct decisions; takeover, interruption response and resume, backchannel response, and responses during Talking to Others therefore measure different policy errors.

\subsection{How to speak: expression, persona, and empathy}
Length, style, character consistency, emotional appropriateness, and spoken naturalness constrain response content. The specialist pipeline uses models trained with SFT and GRPO as teachers. Session-level rollouts score repetition, templated follow-up questions, dialogue stagnation, and long-range persona drift, while deterministic filters catch truncation, extreme length, repeated openings, and abnormal character patterns. The reward is length-neutral and does not reward persona-keyword matching. Audio-dependent empathy trains the model to use conversationally actionable cues---for example, hesitation, fatigue, or crying---without mechanically restating every audible attribute.

Multi-scene dialogue evaluation separates information query, skill control, instruction and style control, and daily chat because they stress different response policies. In VoiceChat, judged by Qwen-Plus, a turn-level Spoken score assesses response naturalness and request advancement, while a rubric measures explicit understanding, usefulness, speakability, and instruction execution. The empathy evaluations use a common high-empathy persona, so they test behavior under that role constraint rather than every possible persona. The reported persona and empathy results use speech-to-text (S2T) evaluation, which covers input understanding and response content. VoiceChat complements these evaluations with multi-scene conversational scores (Table~\ref{tab:spoken-real}).

\paragraph{Policy-to-evidence map.}
The three policy dimensions have different evidence coverage. ``Whether'' is evaluated through tool decisions in Table~\ref{tab:agent-bench}, reliability and safety tests in Table~\ref{tab:safety-source}, fine-grained safety (Table~\ref{tab:fine-safety-redteam}), and a 50-session human red-team study. ``When'' is evaluated through pause handling, turn-taking, interruption, backchannel, Talking to Others, multi-party interaction, and semantic control in Tables~\ref{tab:fdb-selected} and~\ref{tab:fdb-v3}. ``How'' is evaluated through persona, empathy, and VoiceChat in Tables~\ref{tab:persona-empathy-html} and~\ref{tab:spoken-real}. Larger human red-team studies and long-horizon executable spoken-task evaluation remain future work.

\suppressfloats[t]
\section{Evaluation: Intelligence, Action, Interaction and Realtime}
\label{sec:evaluation}
\subsection{Protocol and evidence scope}
Results should be interpreted under the evaluation setting stated for each table, including the input modality, data, judge, and sample count when available. S2T denotes speech input with text output, S2S denotes speech input with speech output, and T2T denotes text input with text output. Unless stated otherwise, evaluations use S2T; the Full-Duplex-Bench (FDB) series and EVA-A use S2S, while the MC column in Table~\ref{tab:core-bench} uses T2T. The $\tau$-Voice results use our half-duplex S2T adaptation and are not directly comparable to the benchmark's official full-duplex S2S protocol. Unless stated otherwise, GPT-Realtime-2 is evaluated with the low-effort setting.

We use percentage points for absolute differences between percentage metrics and explicitly label relative changes; differences on other scales are score points. A down arrow marks metrics where lower is better; captions and accompanying text define the remaining metric directions and units. Boldface marks the best displayed result for each metric under its stated comparison, including ties; it does not imply statistical significance. A dash denotes an unavailable result. Release-level comparisons do not isolate individual training components. Reported comparisons are descriptive.

Benchmarks with a cited public release use their published names. The in-house evaluations used here are the multilingual BBA extension, Long-AMC, WebSearch1K, PersonaCross, 20+ Turns with Plot Advancement (PA), the Chinese real and two synthetic empathy sets, VoiceChat, VoiceChat-L, Qwen-Audio-FDB, the Chinese and English multi-turn attack sets, and capability boundaries. Their table captions report the available task scope and metric definitions.

\subsection{Intelligence: general, audio, and long-context capabilities}

Tables~\ref{tab:core-bench}--\ref{tab:long-bench} compare Qwen-Audio-3.0-Realtime and Qwen-Audio-3.1-Realtime with representative baselines. OpenAudioBench (OAB) was introduced with Baichuan-Audio~\citep{li2025baichuanaudio}, while VoiceBench~\citep{chen2024voicebench} evaluates voice assistants. The displayed OAB and VoiceBench ``Overall'' values average the scored subitems and may not represent complete official benchmark totals. Big Bench Audio (BBA)~\citep{hill-smith2024bba} and Audio MultiChallenge (AMC)~\citep{gosai2025audiomultichallenge} test audio reasoning and instruction following; the MultiChallenge (MC)~\citep{sirdeshmukh2025multichallenge} results use text input and text output to contextualize multi-turn instruction following. BBA and AMC are percentages; OAB, VoiceBench, and MC use their displayed benchmark scores.

\begin{table*}[htbp]
\centering\tablefont
\setlength{\tabcolsep}{8pt}
\begin{tabular}{lccccc}
\toprule
Model & \metricup{OAB} & \metricup{VoiceBench} & \metricup{BBA} & \metricup{AMC} & \metricup{MC (T2T)} \\
\midrule
GPT-Realtime-2 & 87.31 & 83.37 & 93.30 & 50.33 & 44.32 \\
SeedDuplex 1.2.6.1 & 85.38 & 87.43 & 80.50 & 41.91 & 49.82 \\
Qwen-Audio-3.0-Realtime & \textbf{88.92} & 92.54 & \textbf{98.80} & 47.12 & 52.38 \\
Qwen-Audio-3.1-Realtime & 88.82 & \textbf{92.73} & 98.50 & \textbf{52.21} & \textbf{53.85} \\
\bottomrule
\end{tabular}
\caption{General and instruction-following results. MC uses text-to-text input/output; other columns use speech input. AMC and MC use GPT-4o-mini as judge. Official AMC uses o4-mini~\citep{gosai2025audiomultichallenge}, unavailable here; our AMC scores are not directly comparable.}
\label{tab:core-bench}
\end{table*}

Big Bench Audio (BBA) contains 1,000 English audio questions adapted from four Big Bench Hard categories \citep{hill-smith2024bba}. We construct an in-house multilingual extension by translating the benchmark into 13 additional languages, yielding a 14-language evaluation spanning Arabic, Chinese, English, major European languages, and languages from East and Southeast Asia. The language codes are ar (Arabic), de (German), en (English), es (Spanish), fr (French), id (Indonesian), it (Italian), ja (Japanese), ko (Korean), pt (Portuguese), ru (Russian), th (Thai), vi (Vietnamese), and zh (Chinese). Table~\ref{tab:bba-multilingual} reports accuracy rounded to one decimal; Avg. is the unweighted mean across the 14 languages, computed before rounding.

\begin{table*}[htbp]
\centering\tablefont
\setlength{\tabcolsep}{3.2pt}
\begin{tabular}{l*{15}{c}}
\toprule
Model & ar & de & en & es & fr & id & it & ja & ko & pt & ru & th & vi & zh & Avg. \\
\midrule
GPT-Realtime-2 & \textbf{83.6} & 79.8 & 93.3 & 84.6 & 79.1 & 86.6 & 81.2 & 79.8 & 87.6 & 80.7 & 82.3 & 74.2 & 75.1 & 81.1 & 82.1 \\
Qwen-Audio-3.0-Realtime & 52.5 & 83.4 & \textbf{98.8} & 91.1 & 83.1 & 92.3 & 87.0 & 89.1 & \textbf{90.4} & 85.7 & 86.3 & 51.5 & 61.5 & \textbf{90.9} & 81.7 \\
Qwen-Audio-3.1-Realtime & 79.1 & \textbf{88.9} & 98.5 & \textbf{93.4} & \textbf{87.4} & \textbf{93.1} & \textbf{89.3} & \textbf{89.3} & 89.3 & \textbf{87.6} & \textbf{88.3} & \textbf{79.0} & \textbf{79.2} & \textbf{90.9} & \textbf{88.1} \\
\bottomrule
\end{tabular}
\caption{Multilingual BBA accuracy (\%).}
\label{tab:bba-multilingual}
\end{table*}

Table~\ref{tab:audio-multilingual} evaluates audio understanding and multilingual automatic speech recognition. We report accuracy (in percent) on Massive Multi-Task Audio Understanding and Reasoning (MMAU)~\citep{sakshi2024mmau}. LibriSpeech test-clean (LS-clean) and test-other (LS-other) report word error rate (WER)~\citep{panayotov2015librispeech}. Few-shot Learning Evaluation of Universal Representations of Speech (FLEURS)~\citep{conneau2022fleurs} reports macro-average WER across the 14 language varieties evaluated here.

\begin{table*}[htbp]
\centering\tablefont
\setlength{\tabcolsep}{7pt}
\begin{tabular}{lrrrr}
\toprule
Model & \metricup{MMAU} & \metricdown{LS-clean} & \metricdown{LS-other} & \metricdown{FLEURS} \\
\midrule
Gemini 3.7 Flash & 77.50 & 3.68 & 6.46 & 4.58 \\
Qwen-Audio-3.0-Realtime & \textbf{82.21} & 1.21 & 2.34 & 9.01 \\
Qwen-Audio-3.1-Realtime & 81.60 & \textbf{1.19} & \textbf{2.31} & \textbf{3.98} \\
\bottomrule
\end{tabular}
\caption{Audio understanding and multilingual automatic speech recognition results.}
\label{tab:audio-multilingual}
\end{table*}

Table~\ref{tab:long-bench} reports LongBench v2~\citep{bai2024longbenchv2}, MiniLongBench~\citep{huang2025minilongbench}, and Long-AMC results as percentages. Long-AMC is an in-house extension following Audio MultiChallenge (AMC)~\citep{gosai2025audiomultichallenge}: it expands AMC's 3--8-turn conversations to 25--45 turns per conversation (median 32) to test instruction following across longer audio dialogues. VoiceChat-L is a distinct in-house set of colloquial multi-turn voice chats, spanning 60--132 turns per conversation (median 64); its 1--5 score probes spoken conversational ability over long histories. The two sets target different behaviors and use different score scales.

\begin{table*}[htbp]
\centering\tablefont
\setlength{\tabcolsep}{5pt}
\begin{tabular}{lcccc}
\toprule
Model & \metricup{LongBench v2} & \metricup{MiniLongBench} & \metricup{Long-AMC} & \metricup{VoiceChat-L} \\
\midrule
Qwen-Audio-3.0-Realtime & 43.14 & 55.70 & 44.44 & \textbf{4.61} \\
Qwen-Audio-3.1-Realtime & \textbf{49.90} & \textbf{60.10} & \textbf{52.48} & 4.58 \\
\bottomrule
\end{tabular}
\caption{Long-context results. In-house Long-AMC tests instruction following over 25--45 turns (median 32; GPT-4o-mini judge); in-house VoiceChat-L tests colloquial spoken dialogue over 60--132 turns (median 64).}
\label{tab:long-bench}
\end{table*}

Relative to Qwen-Audio-3.0-Realtime, Qwen-Audio-3.1-Realtime raises AMC from 47.12 to 52.21, while the T2T MC score increases from 52.38 to 53.85. Its multilingual BBA macro-average rises by 6.4 percentage points to 88.1, with the largest absolute gains on Arabic, Thai, and Vietnamese. On speech recognition, FLEURS macro-average WER falls from 9.01 to 3.98, and LibriSpeech clean/other WER reaches 1.19/2.31. LongBench v2, MiniLongBench, and Long-AMC improve by 6.76, 4.40, and 8.04 percentage points, respectively; VoiceChat-L remains close at 4.58 versus 4.61.

\FloatBarrier
\subsection{Action: tool use and retrieval}

Table~\ref{tab:agent-bench} reports speech-conditioned tool use and agent evaluation. $\tau$-Voice~\citep{ray2026tauvoice} covers grounded tasks in the Retail, Airline, and Telecom domains. The results reported here use a half-duplex S2T adaptation rather than the benchmark's official full-duplex S2S protocol. Overall is the reported task-success aggregate, not an unweighted average of the displayed domain scores. SpeechFCEval is the speech function-calling benchmark~\citep{speechfceval,funaudiochat}, and EVA-A is the Accuracy composite metric of EVA-Bench~\citep{bogavelli2026evabench}. Our EVA-A setup differs slightly from the public configuration: audio judging mainly uses Gemini-3-flash-preview, with some supplementary Gemini-3.5-flash scores. The 3.1 result covers 200 of 213 sessions; the remaining sessions were excluded after inference failures or repeated scoring timeouts.

Table~\ref{tab:websearch-date} isolates retrieval triggering and query behavior on WebSearch1K, an in-house S2T evaluation with 1,000 single-turn examples, including 99 positives. It tests whether to call web search and which queries to emit; it does not execute retrieval or score the final grounded answer. Call rate and F1 are percentages, and a 10--20\% call rate with approximately 60\% F1 defines the prespecified operating range. Query counts are summarized by their mean, standard deviation (SD), and range. They describe proposed query volume and redundancy, not measured retrieval cost, answer quality, or latency.

\begin{table*}[htbp]
\centering\tablefont
\setlength{\tabcolsep}{3pt}
\renewcommand{\arraystretch}{1.08}
\begin{tabular}{l*{7}{c}}
\toprule
\multirow[c]{2}{*}{Model} & \multicolumn{4}{c}{$\tau$-Voice (half-duplex)} & \multirow[c]{2}{*}{SpeechFCEval} & \multicolumn{2}{c}{EVA-A} \\
\cmidrule(lr){2-5}\cmidrule(lr){7-8}
& Retail & Airline & Telecom & Overall & & Pass & Mean \\
\midrule
GPT-Realtime-2 & 43.0 & 56.0 & 32.5 & 43.8 & 70.74 & \textbf{50.40} & 68.30 \\
SeedDuplex 1.2.6.1 & 8.8 & 57.1 & 54.9 & 40.3 & 54.73 & -- & -- \\
\midrule
Qwen-Audio-3.0-Realtime & 72.8 & 64.0 & 90.4 & 78.4 & 83.28 & 43.10 & \textbf{70.50} \\
Qwen-Audio-3.1-Realtime & \textbf{73.7} & \textbf{74.0} & \textbf{93.9} & \textbf{82.0} & \textbf{86.00} & 47.74 & 66.26 \\
\bottomrule
\end{tabular}
\caption{Agentic and tool-use results (\%).}
\label{tab:agent-bench}
\end{table*}

\begin{table*}[htbp]
\centering\tablefont
\setlength{\tabcolsep}{6pt}
\begin{tabular}{lccc}
\toprule
Model & Call rate & \metricup{F1} & Queries: mean $\pm$ SD [range] \\
\midrule
GPT-Realtime-2 & 10.10 & 60.00 & 3.19 $\pm$ 1.01 [2--9] \\
SeedDuplex 1.2.6.1 & 93.60 & 18.94 & 1.19 $\pm$ 0.53 [0--9] \\
\midrule
Qwen-Audio-3.0-Realtime & 15.40 & \textbf{60.87} & 4.37 $\pm$ 0.91 [1--9] \\
Qwen-Audio-3.1-Realtime & 17.40 & 58.61 & 1.05 $\pm$ 0.29 [1--4] \\
\bottomrule
\end{tabular}
\caption{In-house WebSearch1K retrieval-triggering and query-generation results.}
\label{tab:websearch-date}
\end{table*}

On the half-duplex S2T $\tau$-Voice adaptation in Table~\ref{tab:agent-bench}, Qwen-Audio-3.1-Realtime raises Overall task success from 78.4\% to 82.0\% relative to 3.0. Retail increases from 72.8\% to 73.7\%, Airline from 64.0\% to 74.0\%, and Telecom from 90.4\% to 93.9\%. These results should not be compared directly with official full-duplex $\tau$-Voice results. SpeechFCEval accuracy increases from 83.28\% to 86.00\%. Under our EVA-A configuration, Qwen-Audio-3.1-Realtime obtains 47.74\% Pass and 66.26\% Mean; 3.0 reports 43.10\% and 70.50\%. This comparison is descriptive because identical session coverage is not established.

\Needspace{8\baselineskip}
On the in-house WebSearch1K setting in Table~\ref{tab:websearch-date}, the mean number of proposed search queries falls from 4.37 to 1.05, a 76.0\% relative reduction. This pattern is consistent with the data-side removal of near-duplicates and the GRPO count penalty in Equation~\ref{eq:query-count-reward}, which lowers the reward when the model emits more queries than the judge estimates are needed. Retrieval-triggering F1 nevertheless decreases from 60.87\% to 58.61\% by 2.26 percentage points; the call rate remains within the target operating range. Since WebSearch1K does not execute retrieval, the query-count reduction does not establish measured savings in search latency or equivalent grounded-answer quality.

\FloatBarrier
\subsection{Interaction: spoken behavior, persona, empathy, and duplex}

Table~\ref{tab:persona-empathy-html} combines persona and empathy evaluations. CharacterEval~\citep{tu2024charactereval} and RMTBench~\citep{xiang2025rmtbench} are public role-playing benchmarks; we use subsets and rescore complete sessions with a spoken role-play rubric. PersonaCross and the 20+ Turns set are in-house: the former crosses diverse personas with spoken conversations, while the latter contains conversations longer than 20 turns and targets character maintenance and Plot Advancement (PA). EchoMind~\citep{zhou2025echomind} is a public empathetic-speech benchmark. Its reported score, and those of the in-house Chinese real, multi-turn synthetic, and single-turn synthetic sets, average Context Fit, Response Naturalness, Colloquialism Degree, and Speech Information Relevance under a common S2T high-empathy persona setting.

\begin{table*}[htbp]
\centering\tablefont
\setlength{\tabcolsep}{2pt}
\renewcommand{\arraystretch}{1.08}
\begin{tabular}{l*{8}{c}}
\toprule
\multirow[c]{2}{*}{Model} & \multicolumn{4}{c}{Persona evaluation} & \multicolumn{4}{c}{Empathy evaluation} \\
\cmidrule(lr){2-5}\cmidrule(lr){6-9}
& CharacterEval & PersonaCross & \multicolumn{1}{c}{\theadcenter{20+ Turns\\(PA)}} & RMTBench & EchoMind & \multicolumn{1}{c}{\theadcenter{Synthetic\\multi-turn}} & \multicolumn{1}{c}{\theadcenter{Synthetic\\single-turn}} & \multicolumn{1}{c}{\theadcenter{In-house\\real}} \\
\midrule
GPT-Realtime-2 & 3.96 & 3.69 & 3.23 & 3.41 & 3.20 & 4.33 & 4.68 & 3.54 \\
SeedDuplex 1.2.6.1 & \textbf{4.18} & 3.57 & \textbf{3.75} & \textbf{3.53} & 3.70 & 4.86 & 4.89 & 4.58 \\
Qwen-Audio-3.0-Realtime & 3.86 & 3.46 & 3.29 & 3.31 & 3.70 & 4.73 & 4.86 & 4.37 \\
Qwen-Audio-3.1-Realtime & 3.93 & \textbf{3.73} & 3.46 & 3.49 & \textbf{4.03} & \textbf{4.99} & \textbf{4.99} & \textbf{4.74} \\
\bottomrule
\end{tabular}
\caption{Persona and empathy results (1--5, higher is better). Final judges: GPT-5.6-luna (persona) and Gemini-3.1-pro-preview (empathy).}
\label{tab:persona-empathy-html}
\end{table*}

VoiceChat is an in-house evaluation of 218 Chinese conversations and 2,706 turns, judged by Qwen-Plus. The displayed scene slices cover Information Query (60 conversations; about 795 turns), Skill Control (45; about 554), Instruction \& Style Control (30; about 315), and Daily Chat (67; about 876); these slices do not exhaust the overall set. The first, second, and fourth slices are grouped by session intent; Instruction \& Style Control is scored per turn because controls can co-occur. Spoken (1--5) measures naturalness and request advancement, while Rubric (0--1) averages 12 binary response-quality criteria. Bold marks the best displayed system score or the better endpoint in a diagnostic comparison. These scores assess conversational performance and should not be interpreted as direct measures of acoustic quality.

\begin{table}[htbp]
\centering\tablefont
\setlength{\tabcolsep}{1.8pt}
\renewcommand{\arraystretch}{1.08}
\begin{tabular}{l*{10}{c}}
\toprule
\multirow[c]{2}{*}{Model} & \multicolumn{2}{c}{Overall} & \multicolumn{2}{c}{\theadcenter{Information\\Query}} & \multicolumn{2}{c}{Skill Control} & \multicolumn{2}{c}{\theadcenter{Instruction \&\\Style Control}} & \multicolumn{2}{c}{Daily Chat} \\
\cmidrule(lr){2-3}\cmidrule(lr){4-5}\cmidrule(lr){6-7}\cmidrule(lr){8-9}\cmidrule(lr){10-11}
& Spoken & Rubric & Spoken & Rubric & Spoken & Rubric & Spoken & Rubric & Spoken & Rubric \\
\midrule
GPT-Realtime-2 & 3.881 & 0.945 & 3.864 & 0.943 & 3.581 & 0.921 & 4.257 & \textbf{0.964} & 3.939 & 0.952 \\
Gemini 3.7 Flash & 3.961 & 0.928 & 4.093 & 0.952 & 3.688 & 0.901 & 4.302 & 0.933 & 3.854 & 0.924 \\
SeedDuplex 1.2.6.1 & 3.270 & 0.844 & 3.208 & 0.844 & 3.042 & 0.802 & 3.727 & 0.863 & 3.340 & 0.868 \\
Qwen-Audio-3.0-Realtime & 4.046 & 0.956 & 4.142 & 0.974 & \textbf{3.951} & \textbf{0.945} & 4.070 & 0.928 & \textbf{4.016} & 0.955 \\
Qwen-Audio-3.1-Realtime & \textbf{4.100} & \textbf{0.961} & \textbf{4.246} & \textbf{0.978} & 3.949 & 0.941 & \textbf{4.380} & 0.939 & 3.965 & \textbf{0.965} \\
\bottomrule
\end{tabular}

\vspace{1pt}
\setlength{\tabcolsep}{9pt}
\begin{tabular}{lcc}
\toprule
Diagnostic slice & Spoken: 3.0 $\rightarrow$ 3.1 & Rubric: 3.0 $\rightarrow$ 3.1 \\
\midrule
Voice/persona switching & 3.469 $\rightarrow$ \textbf{4.106} & 0.890 $\rightarrow$ \textbf{0.910} \\
Composite-emotion switching & 2.929 $\rightarrow$ \textbf{4.000} & 0.845 $\rightarrow$ \textbf{0.905} \\
Daily Chat persona interaction & \textbf{3.685} $\rightarrow$ 3.371 & 0.916 $\rightarrow$ \textbf{0.951} \\
\bottomrule
\end{tabular}

\vspace{1pt}
\setlength{\tabcolsep}{5.2pt}
\begin{tabular}{lrrrrr}
\toprule
Model & Overall & Information Query & Skill Control & \multicolumn{1}{c}{\theadcenter{Instruction \&\\Style Control}} & Daily Chat \\
\midrule
GPT-Realtime-2 & 177 & 227 & 137 & 100 & 183 \\
Gemini 3.7 Flash & 149 & 219 & 109 & 96 & 127 \\
SeedDuplex 1.2.6.1 & 61 & 87 & 55 & 50 & 46 \\
Qwen-Audio-3.0-Realtime & 58 & 71 & 45 & 37 & 58 \\
Qwen-Audio-3.1-Realtime & 53 & 72 & 36 & 38 & 48 \\
\bottomrule
\end{tabular}
\caption{In-house VoiceChat results. The panels report scene-level Spoken/Rubric scores, diagnostic comparisons, and average Chinese characters per assistant turn; response length is descriptive.}
\label{tab:spoken-real}
\end{table}

Relative to Qwen-Audio-3.0-Realtime, Table~\ref{tab:persona-empathy-html} shows that 3.1 raises CharacterEval from 3.86 to 3.93, PersonaCross from 3.46 to 3.73, 20+ Turns (PA) from 3.29 to 3.46, and RMTBench from 3.31 to 3.49. EchoMind rises from 3.70 to 4.03; both synthetic empathy scores reach 4.99, and the real-conversation score reaches 4.74. Across the displayed systems, SeedDuplex leads CharacterEval, 20+ Turns, and RMTBench, while 3.1 leads PersonaCross and all four empathy columns.

On VoiceChat, Qwen-Audio-3.1-Realtime records the highest displayed overall Spoken and Rubric scores at 4.100 and 0.961, gains of 0.054 and 0.005 over 3.0. Information Query improves by 0.104 on Spoken and 0.004 on Rubric. Instruction \& Style Control shows the largest aggregate Spoken gain, from 4.070 to 4.380; voice/persona and composite-emotion switching rise from 3.469 to 4.106 and from 2.929 to 4.000. For Daily Chat, Spoken decreases from 4.016 to 3.965 while Rubric increases from 0.955 to 0.965. Its persona-interaction slice follows the same pattern: Spoken decreases from 3.685 to 3.371 while Rubric increases from 0.916 to 0.951. Skill Control remains close, with 3.1 at 3.949 Spoken and 0.941 Rubric versus 3.951 and 0.945 for 3.0. These results show why Spoken and Rubric should be read together. Mean response length decreases from 58 to 53 Chinese characters per turn, most clearly on Skill Control (45 to 36) and Daily Chat (58 to 48); Information Query and Instruction \& Style Control remain nearly unchanged. Response length is descriptive, and shorter responses are not necessarily better.

Full-Duplex-Bench (FDB) v1.0/v1.5~\citep{lin2025fullduplexbench} covers Pause Handling, Smooth Turn-Taking, User Interruption, User Backchannel, Background Speech, and Talking to Others (silence when the primary user addresses someone else). Takeover rate (TOR) is lower-is-better for Synthetic and Candor Pause Handling but higher-is-better for Candor Smooth Turn-Taking. Table~\ref{tab:fdb-selected} reports rates and latency in seconds; v1.5 includes Respond, Resume, Uncertain, and Unknown for four tasks, plus stop and response latency for interruption and backchannel. FDB v3.0~\citep{lin2026fdbv3} adds tool selection (ToolSel), argument accuracy (ArgAcc), response quality (RespQual), first-attempt success (Pass@1), turn-taking, interruption, and filler behavior.

\Needspace{4\baselineskip}
In-house Qwen-Audio-FDB has 100 sessions in each of two subsets, Multi-party Interaction and Semantic Control, with 700 and 299 judgeable points. A session passes only if all points pass; point-level scores exclude unjudgeable points. Table~\ref{tab:fdb-v3} reports proportions on a 0--1 scale.

\begin{table}[H]
\centering\tablefont
\setlength{\tabcolsep}{4pt}
\renewcommand{\arraystretch}{1.12}
\begin{tabular}{lllrrrr}
\toprule
Benchmark & Task & Metric & \multicolumn{1}{c}{GPT-Realtime-2} & \multicolumn{1}{c}{\theadcenter{SeedDuplex\\1.2.6.1}} & \multicolumn{1}{c}{\theadcenter{Qwen-Audio-\\3.0-Realtime}} & \multicolumn{1}{c}{\theadcenter{Qwen-Audio-\\3.1-Realtime}} \\
\midrule
\multirow[c]{4}{*}[-\dimexpr(\aboverulesep+\belowrulesep+\cmidrulewidth)/2\relax]{FDB v1.0} & \multirow[c]{2}{*}{Pause Handling} & \metricdown{Synthetic TOR} & 0.0146 & 0.0511 & \textbf{0.0073} & \textbf{0.0073} \\
& & \metricdown{Candor TOR} & \textbf{0.0370} & 0.1296 & 0.1100 & 0.0509 \\
\cmidrule(lr){2-7}
& \multirow[c]{2}{*}{Smooth Turn-Taking} & Candor TOR & \textbf{1.0000} & 0.9412 & 0.9580 & 0.9664 \\
& & \metricdown{Latency (s)} & 1.7990 & 1.8440 & \textbf{1.5390} & 1.9210 \\
\midrule
\multirow[c]{20}{*}[-\dimexpr(\aboverulesep+\belowrulesep+\cmidrulewidth)*3/2\relax]{FDB v1.5} & \multirow[c]{6}{*}{User Interruption} & Respond & 0.8350 & 0.6800 & \textbf{0.8800} & 0.8450 \\
& & \metricdown{Resume} & 0.0700 & 0.1850 & \textbf{0.0350} & 0.1300 \\
& & \metricdown{Uncertain} & 0.0200 & 0.0350 & 0.0100 & \textbf{0.0050} \\
& & \metricdown{Unknown} & 0.0750 & 0.1000 & 0.0750 & \textbf{0.0200} \\
& & \metricdown{Stop latency (s)} & \textbf{0.3830} & 1.4190 & 1.0410 & 1.1160 \\
& & \metricdown{Response latency (s)} & \textbf{1.8720} & 2.1080 & 1.9432 & 1.9540 \\
\cmidrule(lr){2-7}
& \multirow[c]{6}{*}{User Backchannel} & \metricdown{Respond} & 0.0204 & \textbf{0.0000} & \textbf{0.0000} & 0.0306 \\
& & Resume & 0.9184 & 0.9796 & \textbf{0.9800} & 0.9694 \\
& & \metricdown{Uncertain} & 0.0102 & \textbf{0.0000} & \textbf{0.0000} & \textbf{0.0000} \\
& & \metricdown{Unknown} & 0.0510 & 0.0204 & 0.0200 & \textbf{0.0000} \\
& & Stop latency (s) & 0.3930 & 0.7240 & \textbf{0.7404} & 0.7380 \\
& & \metricdown{Response latency (s)} & 1.9130 & 2.1430 & 1.8345 & \textbf{1.7000} \\
\cmidrule(lr){2-7}
& \multirow[c]{4}{*}{Background Speech} & \metricdown{Respond} & 0.7200 & 0.6100 & 0.7300 & \textbf{0.1300} \\
& & Resume & 0.1200 & 0.2900 & 0.2600 & \textbf{0.8700} \\
& & \metricdown{Uncertain} & 0.0600 & 0.0300 & \textbf{0.0000} & \textbf{0.0000} \\
& & \metricdown{Unknown} & 0.1000 & 0.0700 & 0.0100 & \textbf{0.0000} \\
\cmidrule(lr){2-7}
& \multirow[c]{4}{*}{Talking to Others} & \metricdown{Respond} & 0.6000 & 0.8100 & 0.1300 & \textbf{0.0300} \\
& & Resume & 0.1300 & 0.0500 & 0.8200 & \textbf{0.9600} \\
& & \metricdown{Uncertain} & 0.0300 & 0.0200 & 0.0100 & \textbf{0.0000} \\
& & \metricdown{Unknown} & 0.2400 & 0.1200 & 0.0400 & \textbf{0.0100} \\
\bottomrule
\end{tabular}
\caption{FDB v1.0 and v1.5 results, grouped by task. Rates are proportions; latency is in seconds. Down arrows mark lower-is-better metrics; other metrics are higher-is-better. Bold indicates the best result for each metric, including ties.}
\label{tab:fdb-selected}
\end{table}

\begin{table}[H]
\centering\tablefont
\setlength{\tabcolsep}{3.5pt}
\renewcommand{\arraystretch}{1.08}
\begin{tabular}{lrrrrrrr}
\toprule
\multirow[c]{2}{*}{Model} & \multicolumn{4}{c}{Tool Use} & \multicolumn{3}{c}{Turn-Taking Dynamics} \\
\cmidrule(lr){2-5}\cmidrule(lr){6-8}
& \multicolumn{1}{c}{\metricup{ToolSel}} & \multicolumn{1}{c}{\metricup{ArgAcc}} & \multicolumn{1}{c}{\metricup{RespQual}} & \multicolumn{1}{c}{\metricup{Pass@1}} & \multicolumn{1}{c}{\metricup{Take-turn}} & \multicolumn{1}{c}{\metricdown{Interrupt}} & \multicolumn{1}{c}{\metricdown{Filler}} \\
\midrule
GPT-Realtime-2 & 0.9610 & 0.6600 & \textbf{0.8800} & 0.5600 & \textbf{1.0000} & 0.1000 & 1.0000 \\
SeedDuplex 1.2.6.1 & \textbf{0.9760} & \textbf{0.7260} & 0.7810 & \textbf{0.6000} & 0.9600 & \textbf{0.0420} & 0.4460 \\
Qwen-Audio-3.0-Realtime & 0.8870 & 0.6400 & 0.5800 & 0.5700 & \textbf{1.0000} & 0.1700 & 0.7590 \\
Qwen-Audio-3.1-Realtime & 0.9370 & 0.6260 & 0.6970 & 0.5400 & 0.9900 & 0.1110 & \textbf{0.2960} \\
\bottomrule
\end{tabular}

\vspace{3pt}
\setlength{\tabcolsep}{8pt}
\begin{tabular}{lrrrr}
\toprule
\multirow[c]{2}{*}{Model} & \multicolumn{2}{c}{Multi-party Interaction} & \multicolumn{2}{c}{Semantic Control} \\
\cmidrule(lr){2-3}\cmidrule(lr){4-5}
& \multicolumn{1}{c}{Session-level} & \multicolumn{1}{c}{Point-level} & \multicolumn{1}{c}{Session-level} & \multicolumn{1}{c}{Point-level} \\
\midrule
GPT-Realtime-2 & 0.0700 & 0.8271 & 0.0500 & 0.5017 \\
SeedDuplex 1.2.6.1 & 0.9400 & 0.9900 & 0.3500 & 0.6856 \\
Qwen-Audio-3.0-Realtime & 0.0800 & 0.7814 & 0.0700 & 0.5284 \\
Qwen-Audio-3.1-Realtime & \textbf{0.9600} & \textbf{0.9943} & \textbf{0.6800} & \textbf{0.8930} \\
\bottomrule
\end{tabular}
\caption{FDB v3.0 (upper panel) and in-house Qwen-Audio-FDB (lower panel) results, grouped by task. All values are reported on a 0--1 scale with four decimal places; rates are expressed as proportions. Down arrows mark lower-is-better metrics; other metrics are higher-is-better.}
\label{tab:fdb-v3}
\end{table}

On FDB v1.0, 3.1 ties 3.0 for the lowest Synthetic pause-handling TOR at 0.0073, while its Candor Pause Handling TOR is lower than 3.0 (0.0509 versus 0.1100). For Smooth Turn-Taking, its Candor TOR is 0.9664 and latency is 1.9210 seconds. On FDB v1.5 Background Speech, the response rate falls from 0.7300 to 0.1300 and the resume rate rises from 0.2600 to 0.8700. On Talking to Others, the response rate falls from 0.1300 to 0.0300 and the resume rate rises from 0.8200 to 0.9600. For User Interruption, 3.1 records Respond/Resume rates of 0.8450/0.1300 and Uncertain/Unknown rates of 0.0050/0.0200. For User Backchannel, its response and resume rates are 0.0306 and 0.9694; both Uncertain and Unknown are 0.0000, and response latency is 1.7000 seconds, the lowest displayed value.

On FDB v3.0, 3.1 reduces the filler rate from 0.7590 to 0.2960 relative to 3.0 while recording 0.9900 take-turn accuracy and 0.1110 interruption error. On the in-house Qwen-Audio-FDB sets, session-level pass rate reaches 0.9600 for multi-party interaction and 0.6800 for semantic control, compared with 0.0800 and 0.0700 for 3.0. Point-level pass rates reach 0.9943 and 0.8930, respectively. The low 3.0 session rates reflect the strict all-points rule: a single reply to the wrong speaker or premature turn fails a session.

\FloatBarrier

\subsection{Reliability and safety}

The public single-turn evaluation uses all 450 Chinese Hallucination Question-Answering (HalluQA) questions~\citep{cheng2023halluqa}, the 790-item TruthfulQA release used in our evaluation~\citep{lin2021truthfulqa}, all 939 Do-Not-Answer (DNA) prompts~\citep{wang2024donotanswer}, and 2,726 released Chinese Do-Not-Answer (CDNA) prompts~\citep{wang2024chinese}. HalluQA and TruthfulQA report no-hallucination pass rates, while DNA and CDNA report safety pass rates. The two in-house multi-turn attack sets contain 200 sessions each and report attack success rate, where lower is better. The in-house capability-boundary evaluation contains 68 conversations and 619 responses and reports response-level accuracy.

\begin{table*}[htbp]
\centering\tablefont
\setlength{\tabcolsep}{5pt}
\renewcommand{\arraystretch}{1.08}
\begin{tabular}{l*{7}{c}}
\toprule
\multirow[c]{2}{*}{Model} & \multicolumn{4}{c}{Public single-turn} & \multicolumn{3}{c}{In-house multi-turn and boundary} \\
\cmidrule(lr){2-5}\cmidrule(lr){6-8}
& \metricup{HalluQA} & \metricup{TruthfulQA} & \metricup{DNA} & \metricup{CDNA} & \multicolumn{1}{c}{\theadcenter{\metricdown{Multi-zh}\\Attack}} & \multicolumn{1}{c}{\theadcenter{\metricdown{Multi-en}\\Attack}} & \metricup{Boundary} \\
\midrule
GPT-Realtime-2 & 63.33 & \textbf{83.92} & 88.60 & 81.62 & 42.00 & 30.00 & 94.18 \\
SeedDuplex 1.2.6.1 & 63.33 & 62.41 & 77.96 & 69.52 & 91.50 & 89.50 & 94.35 \\
Qwen-Audio-3.0-Realtime & 68.44 & 78.73 & 91.80 & 83.49 & 80.50 & 63.50 & 94.51 \\
Qwen-Audio-3.1-Realtime & \textbf{71.33} & 82.28 & \textbf{94.99} & \textbf{90.57} & \textbf{26.00} & \textbf{23.50} & \textbf{96.77} \\
\bottomrule
\end{tabular}
\caption{S2T safety and reliability results (\%).}
\label{tab:safety-source}
\end{table*}

Table~\ref{tab:safety-source} reports the available reliability and safety results. The HalluQA and TruthfulQA pass rates combine factuality and refusal behavior under their respective task protocols. Compared with Qwen-Audio-3.0-Realtime, Qwen-Audio-3.1-Realtime gains 2.89, 3.55, 3.19, 7.08, and 2.26 percentage points on HalluQA, TruthfulQA, DNA, CDNA, and capability-boundary accuracy. Its multi-turn attack success rate falls by 54.50 percentage points in Chinese and 40.00 percentage points in English. Across the displayed systems, 3.1 leads six of the seven metrics; GPT-Realtime-2 has the highest TruthfulQA score at 83.92\%. These benchmark results do not measure authorization guarantees enforced by the runtime.

We further aggregate the single- and multi-turn safety cases into ten risk categories. The category-level safety pass rate measures whether the response follows the expected safe behavior for that request and context. Table~\ref{tab:fine-safety-redteam} shows that Qwen-Audio-3.1-Realtime scores above GPT-Realtime-2 in nine of the ten categories. The largest displayed gains are on anthropomorphism and emotional dependency (14.93 percentage points), hate and abusive content (14.19 points), false or misleading information (11.77 points), and mental health crises (10.94 points). Adult content is the only category in which its score is lower, by 0.90 points.

\begin{table*}[htbp]
\centering\tablefont
\setlength{\tabcolsep}{9pt}
\renewcommand{\arraystretch}{1.06}
\begin{tabular}{lcc}
\toprule
Risk category & GPT-Realtime-2 & Qwen-Audio-3.1-Realtime \\
\midrule
Adult content & \textbf{91.89} & 90.99 \\
Assistance with illegal activities & 87.79 & \textbf{95.15} \\
Mental health crises & 74.72 & \textbf{85.66} \\
Privacy and sensitive information & 87.45 & \textbf{92.68} \\
Anthropomorphism and emotional dependency & 71.64 & \textbf{86.57} \\
Violence and physical harm & 81.36 & \textbf{84.75} \\
Hate and abusive content & 79.24 & \textbf{93.43} \\
Discrimination and bias & 84.35 & \textbf{89.57} \\
Fraud and malicious manipulation & 81.97 & \textbf{92.68} \\
False or misleading information & 71.90 & \textbf{83.67} \\
\bottomrule
\end{tabular}
\caption{Fine-grained safety pass rates (\%, higher is better), aggregating single- and multi-turn cases by risk category.}
\label{tab:fine-safety-redteam}
\end{table*}

\paragraph{Human red-team interactions.}
Five testers conduct 50 multi-turn sessions, one per tester per risk category, adapting follow-up questions and attack strategies to the model's replies. An automatic judge scores the session histories: Qwen-Audio-3.1-Realtime achieves a session-level safety pass rate of 92.00\%, compared with 96.00\% for GPT-Realtime-2. Under repeated questioning and targeted prompting, 3.1 maintains rubric-compliant safety behavior in most sessions, although some safety failures remain.

\FloatBarrier

\section{From Realtime Model to Persistent Voice Agent}
Real-time dialogue becomes a persistent voice agent when the interaction loop can hand off durable work, recover task state across turns, and reuse bounded user context. The foreground--background separation in our system is informed by the open-source Qwen-Audio-Agent runtime~\citep{qwenaudioagent2026}. The Voice Harness described here combines that separation with explicit task-state and bounded-memory interfaces. These system mechanisms complement the model capabilities described above while preserving a clear boundary between conversational behavior, executable state, and stored information.

\subsection{Foreground--background coordination}
\begin{figure*}[!htbp]
  \centering
  \includegraphics[width=\textwidth]{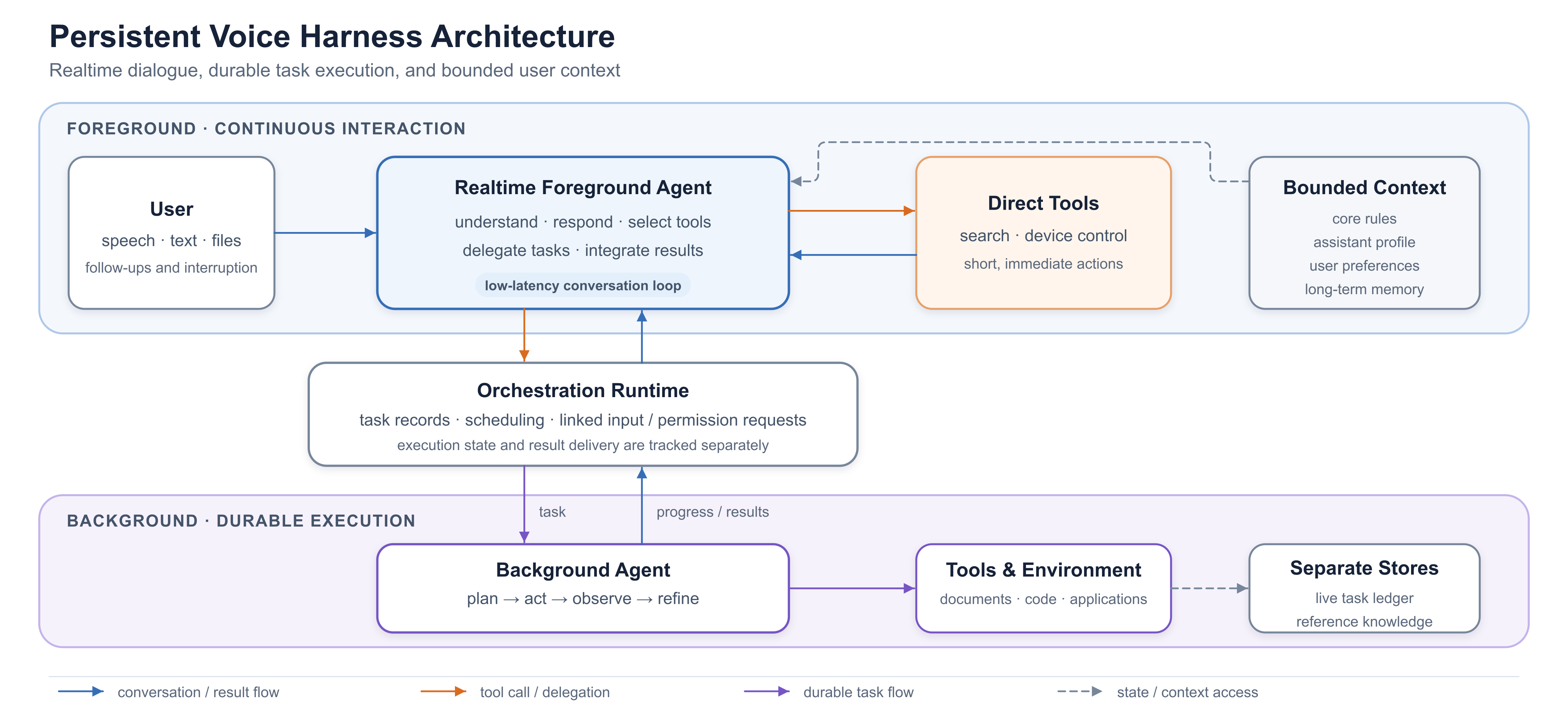}
  \caption{Persistent Voice Harness architecture. The foreground handles dialogue and direct tools. The background agent returns progress and results to the orchestration runtime for delivery to the foreground. Memory, task state, and reference knowledge remain separate. This system is evaluated separately from Qwen-Audio-3.1-Realtime.}
  \label{fig:persistent-agent-architecture}
\end{figure*}

Figure~\ref{fig:persistent-agent-architecture} shows the system organization. The real-time foreground interprets streaming input, selects tools, maintains the dialogue, and expresses results. It can execute short actions directly or submit a self-contained objective, explicit constraints, and referenced inputs to an orchestration runtime. The runtime creates a durable task record and returns an acceptance event so that the foreground can continue serving the conversation. A background agent then performs multi-step work against its own tools and environment, while progress, follow-up questions, permission requests, and terminal results remain linked to the originating task.

This division follows task requirements rather than a fixed number of tool calls. An immediate device operation can remain in the foreground, whereas document processing, code execution, or environment exploration can be delegated. Background executors can be replaced through adapters without changing the foreground interaction contract. Their internal models, tools, and sessions remain executor-managed.

\subsection{Task lifecycle and result delivery}
An acceptance event records successful submission; it does not assert that execution has begun or completed. Tasks progress through queued and running states before reaching completed, failed, or cancelled. Executor-specific delegated or finalizing stages can refine this lifecycle without changing the foreground contract. Follow-up inputs and permission requests attach to the task record instead of becoming independent tasks.

\begin{figure*}[!htbp]
  \centering
  \includegraphics[width=\textwidth]{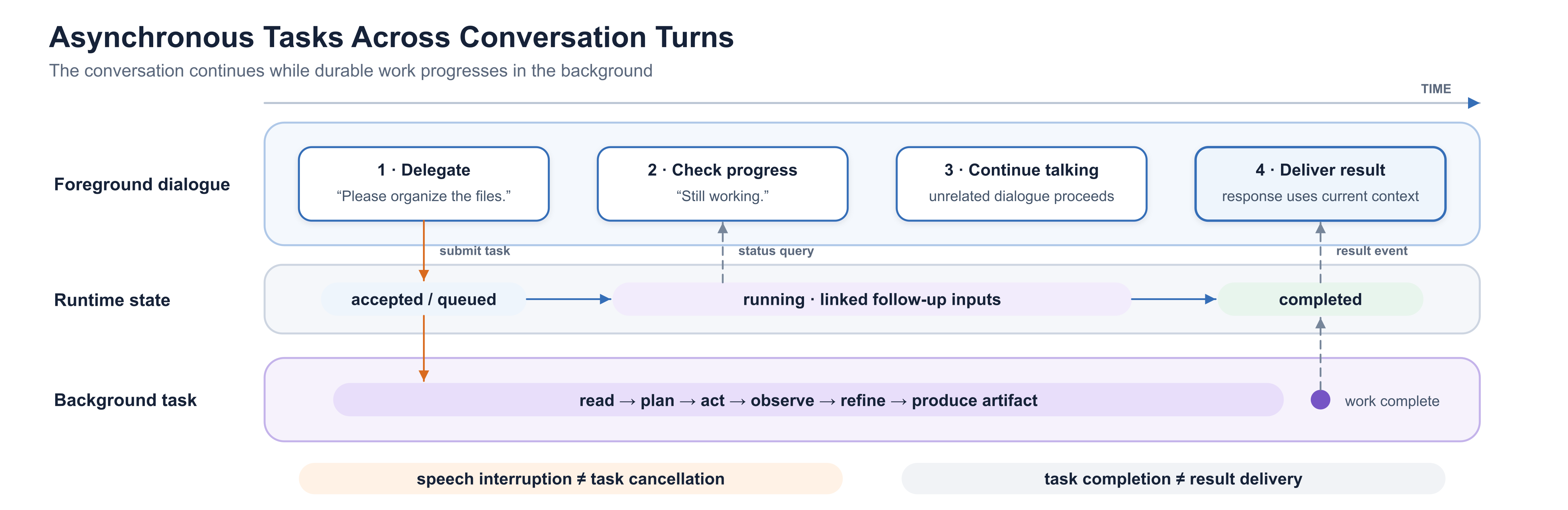}
  \caption{Asynchronous task execution across foreground conversation turns. The runtime keeps task state and linked inputs while dialogue continues. Work completion emits a result event that the foreground incorporates using the current context; speech interruption, task cancellation, task completion, and result delivery retain separate semantics. Horizontal position is illustrative and does not represent measured latency.}
  \label{fig:persistent-agent-timeline}
\end{figure*}

Figure~\ref{fig:persistent-agent-timeline} illustrates how one background task can span multiple speech turns. During execution, the user can request progress, supply additional information, or discuss another topic. Speech interruption stops the current response, while task cancellation requires an explicit control event and confirmation from the executor. Execution completion and result delivery are also recorded separately: a completed artifact can wait for an appropriate conversational moment, and interrupted playback does not invalidate the completed work.

\subsection{System-level routing evaluation}
We separately evaluate the routing design on an in-house cockpit task set. The evaluated prototype uses Qwen-Audio-3.0-Realtime as the real-time audio foreground and a text background executor; it is not an evaluation of the released Qwen-Audio-3.1-Realtime model. The capability split contains 134 cases: 86 short or context-dependent interactions and 48 multi-step tasks with planning constraints and environment feedback. All execution paths share tool definitions, initial state, and scoring rules. Audio paths receive the same synthesized speech, while the text-background baseline receives the corresponding text.

Tool F1 and argument accuracy are macro-averaged against fixed reference calls. Task success requires the complete execution trace and business state to satisfy the goal, constraints, and safety conditions; the four rows contain 97, 108, 121, and 122 successful cases out of 134. Reply quality is a binary text judgment. Latency is measured on a separate common-success subset of 71 cases and 80 tool-request turns: 65 short cases produce 74 turns, and six multi-step cases produce six turns. Audio latency begins when user speech ends, whereas text latency begins at request submission. Both end when the necessary tool result first satisfies the turn objective. Summary generation and speech playback are excluded.

\begin{table*}[htbp]
\centering\tablefont
\setlength{\tabcolsep}{5pt}
\renewcommand{\arraystretch}{1.08}
\begin{tabular}{lcccc}
\toprule
\multirow[c]{2}{*}{Execution path} & \multicolumn{4}{c}{Capability evaluation (\%)} \\
\cmidrule(lr){2-5}
& \metricup{Tool F1} & \metricup{Argument accuracy} & \metricup{Task success} & \metricup{Reply quality} \\
\midrule
Audio foreground, direct tools & 94.7 & 93.6 & 72.39 & 95.5 \\
Audio foreground, all delegated & 90.8 & 89.8 & 80.60 & 89.6 \\
Text background, direct tools & \textbf{96.3} & 94.1 & 90.30 & \textbf{100.0} \\
Audio hybrid routing & 96.2 & \textbf{95.2} & \textbf{91.04} & 97.8 \\
\bottomrule
\end{tabular}

\vspace{4pt}
\begin{tabular}{lccc}
\toprule
\multirow[c]{2}{*}{Execution path} & \multicolumn{3}{c}{Latency on common-success subset (s)} \\
\cmidrule(lr){2-4}
& \metricdown{Short tasks} (74 turns) & \metricdown{Multi-step tasks} (6 turns) & \metricdown{Overall} (80 turns) \\
\midrule
Audio foreground, direct tools & \textbf{1.488} & 67.705 & 6.455 \\
Audio foreground, all delegated & 4.266 & 38.653 & 6.845 \\
Text background, direct tools & 3.149 & \textbf{35.449} & 5.572 \\
Audio hybrid routing & 1.591 & 43.430 & \textbf{4.729} \\
\bottomrule
\end{tabular}
\caption{In-house system-level routing evaluation. The upper panel reports capability percentages on 134 cockpit cases; the lower panel reports mean seconds on a separate common-success subset. Bold marks the highest displayed capability score or lowest displayed latency in each column. The prototype uses Qwen-Audio-3.0-Realtime as its audio foreground and is not a Qwen-Audio-3.1-Realtime model evaluation.}
\label{tab:routing-system-eval}
\end{table*}

Table~\ref{tab:routing-system-eval} shows that hybrid routing reaches 91.04\% task success. From the exact success counts, this is 18.66 percentage points above foreground-only execution and 10.45 percentage points above delegating every audio request. It remains close to the 90.30\% text-background result while retaining audio input. On the separate latency subset, hybrid routing obtains the lowest overall mean at 4.729 seconds: all 65 short cases (74 request turns) remain in the foreground, while all six multi-step cases (six request turns) are delegated. Short-task latency is therefore close to foreground-only execution (1.591 versus 1.488 seconds), and multi-step latency is lower than foreground-only execution (43.430 versus 67.705 seconds).

\Needspace{9\baselineskip}
These results support complementarity between direct and delegated execution under this protocol. They do not measure the formal asynchronous lifecycle in Figures~\ref{fig:persistent-agent-architecture} and~\ref{fig:persistent-agent-timeline}: the benchmark uses synthesized speech, simulated business tools, and a direct delegation interface. Capability and latency come from different batches, and common-success filtering conditions the latency comparison on all four paths succeeding rather than estimating full-set latency. The multi-step latency subset contains six turns, no repeated-seed significance test is reported, and the reply judge is the same model family used by the background executor.

\subsection{Memory and profile}
The foreground context is divided into four layers: core interaction rules, an assistant profile, explicit user preferences, and long-term user memory. Conflicts are resolved by a fixed precedence order: core rules, the user's current explicit request, stored preferences, and the assistant profile. Long-term memory supplies factual context outside this instruction order, and the user's current statement supersedes an older memory. This structure prevents persona or remembered content from overriding tool, permission, or safety rules.

User preferences and memory are bounded logical documents behind a replaceable provider. The lightweight provider maintains a small readable snapshot and exposes atomic \texttt{read}, \texttt{append}, and \texttt{replace} operations. Writes pass through the runtime, use revision checks to avoid silent overwrites, and reject credential-like content. Remote or semantic providers may add retrieval, but the prompt path reads a bounded local snapshot and does not wait for remote I/O\@. Audio observation on the streaming path is restricted to bounded in-memory work; indexing, model calls, and persistence occur at session boundaries.

Personal memory, reference knowledge, and live task state serve different purposes. Memory captures stable user information, a knowledge provider resolves user-supplied reference material, and the task ledger remains the source of truth for changing execution status. Optional profile learning runs after a session and updates a small set of user attributes only after corroborating evidence from multiple sessions. Explicit preferences and inferred attributes remain physically separated, and promoted updates take effect in the next session.

\section{Limitations and Safety}
Evidence coverage remains uneven across the three policy dimensions. The reported 3.1 results cover persona, empathy, VoiceChat, full-duplex behavior, tool use, retrieval triggering and query generation, EVA-A, S2T safety, and a 50-session human red-team study. The duplex evaluation combines public benchmarks with controlled in-house sets; broader real-recording and deployment-condition coverage remains needed. The EVA-A result covers 200 of 213 sessions under an evaluation configuration that differs slightly from the official setup. Long-AMC probes instruction following, VoiceChat-L probes colloquial spoken dialogue, and 20+ Turns probes persona persistence, but the suite does not evaluate long-horizon executable spoken tasks or large-scale human red-teaming.

Most comparisons are descriptive and use public benchmarks or documented in-house sets. Some use automatic judges, synthesized speech, or small slices; suite-wide confidence intervals, repeated-seed tests, and contamination analyses are unavailable. Controlled ablations are needed to attribute changes to Core-Cocktail SFT, either OPD path, environment evolution, or rollout granularity.

The persistent-agent evidence comes from a separate Voice Harness and simulated routing evaluation whose audio foreground is Qwen-Audio-3.0-Realtime. It supports foreground--background routing under that protocol, leaving end-to-end evaluation of the released 3.1 model and formal asynchronous lifecycle open. Runtime memory, permission, and task-state mechanisms require validation beyond learned model compliance.

Future evaluation should expand duplex testing to real recordings and deployment conditions, and add long-horizon executable tasks, larger human safety studies, and end-to-end latency and cost measurements.

\section{Conclusion}
\model{} connects spoken intelligence, executable actions, and conversational policy through Think, Act, and Speak and Coordinate. Foundation post-training transfers language capabilities and consolidates audio-domain expertise. At the center of action learning, self-evolving executable environments and multi-granularity rollouts train task completion and policy compliance. Conversational policy governs whether, when, and how the assistant speaks or acts.

Under the reported settings, 3.1 improves on 3.0 in audio instruction following, multilingual reasoning, speech-conditioned tool use, and selected conversational and full-duplex behaviors. Public and in-house safety evaluations complement these capability measurements. A separate Voice Harness design adds task persistence and bounded memory; its routing study uses a 3.0 foreground and remains distinct from the 3.1 model evaluation. Further work should test these capabilities together in long-horizon spoken tasks with real recordings and end-to-end latency measurements.
\Needspace{9\baselineskip}
\section{Contributions and Acknowledgments}
All authors of Qwen-Audio-3.1-Realtime are listed in alphabetical order by last name.

\noindent\textbf{Authors:} Lujia Bao, Qian Chen, Luyao Cheng, Chong Deng, Yuxiang Kong, Xiangang Li, Xu Li, Jiaqing Liu, Chao-Hong Tan, Haoyu Wang, Wen Wang, Xilou Wang, Haoxiang Xu, Junhao Xu, Liang Yi, Binbin Zhang, Qinglin Zhang, and Qiquan Zhang.

\small
\setlength{\bibsep}{0pt}
\bibliography{references}
\bibliographystyle{colm2026_conference}
\clearpage
\appendix
\fontencoding{T1}\selectfont
\normalsize
\section{Anatomy of an Executable Environment}
\label{app:env-anatomy}

This appendix illustrates the domain and task definitions in Section~\ref{sec:act:build}. The example domain, \texttt{flight\_search\_desk}, supports flight search, date-grid pricing, airport lookup, and refundable \emph{fare holds}. It cannot issue tickets or take payment. The environment bundles a natural-language business policy, a tool pool, and a database (Listings~\ref{lst:app-policy}--\ref{lst:app-db}); a task is a single episode drawn from it (Listing~\ref{lst:app-task}).

\paragraph{Business policy.}
The policy is a natural-language document that enters the agent's system prompt. It states positive capabilities (search, pricing, holds) and, just as importantly, negative constraints: no ticketing, no payment handling, only \texttt{flex} fares are holdable, and at most two active holds per customer. These constraints become task requirements and behavioral assertions; Listing~\ref{lst:app-task} illustrates the checks for one fare-hold request.

\begin{lstlisting}[caption={Business policy of \texttt{flight\_search\_desk} (excerpt, abridged).},label={lst:app-policy}]
You answer calls for the flight search desk: one-way and round-trip
flight search, date-grid pricing, airport lookup, and refundable fare
holds.
Service scope
- Tickets are NOT issued on this desk: booking, payment and ticket
  issuance must be done on the airline website or app. This desk has
  no payment capability: never ask for, accept or write down card or
  payment details, and never pass them on to anyone else.
- The search supports one-way and round-trip trips only. A multi-city
  itinerary cannot be searched as one trip; offer to search each leg
  separately as a one-way trip.
Fare holds
- Only `flex` fares can be held. `saver` and `standard` fares cannot
  be held; if the customer asks, refuse, explain that only flex
  (refundable) fares are holdable, and offer the flex fare of the
  same flight instead.
- A customer may have at most 2 active fare holds. A hold counts as
  active when its `status` is `active` AND its `expiry_date` is later
  than today; expired holds do not count toward the limit, regardless of their
  stored status.
- A hold lasts 72 hours: it is created today and its `expiry_date` is
  3 days later.
\end{lstlisting}

\paragraph{Tool pool.}
Tools are Python functions over the episode's state (Listing~\ref{lst:app-tools}). They follow two conventions. Every tool takes the episode's private database copy as its first argument, so all effects are visible to the post-episode checks. The JSON schema shown to the model is derived mechanically from the signature, type annotations, and docstring, keeping the advertised interface tied to the executable function. Failures are returned as data (\texttt{\{"error": ...\}}) rather than raised, letting the agent recover within the episode.

\begin{lstlisting}[caption={One tool from the pool (body abridged).},label={lst:app-tools},language=Python]
def place_hold(db: dict, phone: str, flight_no: str, depart_date: str,
               fare_class: FARE_CLASSES) -> dict:
    """Place a fare hold for the customer, locking the quoted fare on
    one flight. The hold is created today, runs for 72 hours, and the
    new hold reference is returned.
    Args:
        phone: phone number registered on the account, a digit string
        flight_no: flight number, e.g. SK2859
        depart_date: flight date, YYYY-MM-DD
        fare_class: fare to hold --- saver, standard or flex
    """
    # resolve the account by phone; validate fare class, flight and
    # remaining seats --- any failure returns {"error": ...} and
    # leaves the state untouched
    hold_ref = _next_hold_ref(db)      # e.g. HD482902
    created  = db["current_date"]
    expiry   = _add_days(created, 3)   # 72-hour hold
    db["holds"][hold_ref] = { ..., "fare_class": fare_class,
        "amount": flight["fares"][fare_class], "status": "active",
        "created_date": created, "expiry_date": expiry }
    return {"hold_ref": hold_ref, "status": "active",
            "amount": flight["fares"][fare_class],
            "created_date": created, "expiry_date": expiry}
\end{lstlisting}

\paragraph{Database state.}
The database is a plain JSON document (Listing~\ref{lst:app-db}). All tasks in the environment share this \emph{schema}, while each episode executes on its own fresh copy of an initial \emph{state} (Section~\ref{sec:act:build}). A \texttt{current\_date} field pins ``today'' so derived quantities such as hold expiry are exactly checkable. The seed state is adversarial by design: the one existing hold has \texttt{status=active} but an \texttt{expiry\_date} already before \texttt{current\_date}, so counting active holds by status alone gives an incorrect total.

\begin{lstlisting}[caption={Initial database state (annotated excerpt; the full file has 16 flights and 4 airports).},label={lst:app-db}]
{
  "current_date": "2026-08-12",
  "users": {
    "U70001": {"user_id": "U70001", "name": "Emily Carter",
               "phone": "4155501234"} },
  "flights": {
    "SK2859@2026-08-20": {
      "flight_no": "SK2859", "date": "2026-08-20",
      "origin": "SFO", "dest": "NRT",
      "fares":      {"saver": 640.0, "standard": 780.0, "flex": 990.0},
      "seats_left": {"saver": 4,     "standard": 9,     "flex": 6} }
    // ... 15 more flights
  },
  "holds": {
    "HD482901": {"user_id": "U70001", "flight_no": "SK2859",
      "date": "2026-08-15", "fare_class": "flex", "amount": 955.0,
      "status": "active", "created_date": "2026-08-05",
      "expiry_date": "2026-08-08"}   // expired vs. current_date
  }
}
\end{lstlisting}

\paragraph{Task tuple.}
Listing~\ref{lst:app-task} shows a task tuple for one episode drawn from this environment. Each field plays a distinct role in the episode contract of Section~\ref{sec:act:build}.

\begin{lstlisting}[caption={One task tuple (abridged).},label={lst:app-task}]
{
  "task_id": "write_hold_flex_basic",
  "db": "@dbs/main.json",
  "user_system_prompt":
    "You are Emily Carter (phone 4155501234). You want to put a fare
     hold on flight SK2859 from San Francisco (SFO) to Tokyo Narita
     (NRT) on August 20 --- you want the refundable Flex fare. You say
     the flight number and the date quite fast, almost in one breath.
     You don't know the exact hold rules. If the assistant reads the
     details back to you, confirm them.",
  "expected_outcome": "write",
  "expected_db": {"holds": {"HD482902": {
     "user_id": "U70001", "flight_no": "SK2859", "date": "2026-08-20",
     "origin": "SFO", "dest": "NRT", "fare_class": "flex",
     "amount": 990.0, "status": "active",
     "created_date": "2026-08-12", "expiry_date": "2026-08-15"}}},
  "reference_tool_calls": [
    {"tool": "search_flights", "arguments": {"origin": "SFO",
       "dest": "NRT", "depart_date": "2026-08-20",
       "trip_type": "one_way"}},
    {"tool": "place_hold", "arguments": {"phone": "4155501234",
       "flight_no": "SK2859", "depart_date": "2026-08-20",
       "fare_class": "flex"}} ],
  "assertions": [
    "Must call search_flights to verify flight SK2859 on 2026-08-20
     still has Flex fare seats before calling place_hold.",
    "Before placing the hold, must read back flight SK2859, the date
     2026-08-20, the Flex fare class, the held amount 990.0 and the
     expiry date to Emily, and get her explicit confirmation.",
    "After the hold is placed, must tell Emily the hold reference
     HD482902 and the expiry date 2026-08-15."]
}
\end{lstlisting}

\begingroup
\interlinepenalty=10000
\begin{itemize}
  \item \texttt{db}: a pointer to the initial-state file, deep-copied for every rollout, so episodes never contaminate each other.
  \item \texttt{user\_system\_prompt}: the script handed to the user simulator---identity, goal, the closed set of facts the user knows, and spoken-style cues (the flight number and date are said fast, ``almost in one breath'') that exercise the audio front end rather than tool logic.
  \item \texttt{expected\_outcome}: \texttt{write}---this episode must end in a state change; refusal and unsupported tasks instead require the state to remain untouched.
  \item \texttt{expected\_db}: a subset constraint on the terminal state---the episode's private final state must contain these values at these paths, while a separate write-set check protects unrelated records. The permitted write set is derived from state changes during reference replay; a write outside this set causes the episode to fail.
  \item \texttt{reference\_tool\_calls}: a reference tool-call chain. The construction checks (Section~\ref{sec:act:build}) replay it from the initial state with no model in the loop and require \texttt{expected\_db} to hold; tasks whose reference cannot produce the expected state are repaired or discarded. Dialogue requirements, including confirmation, are checked separately during rollout scoring.
  \item \texttt{assertions}: natural-language behavioral checks, scored after the episode. They check requirements that database state alone cannot capture: search before hold, read-back with explicit confirmation before writing, and truthful reporting of the new hold reference and expiry after writing.
\end{itemize}
\endgroup

Read together, the four listings also show where each check sits in time: construction checks replay the reference solution from the initial state against the expected outcome; during the episode, the environment executes tools truthfully against the private state; and after the episode, the terminal-state subset check, the derived write-set veto, and the assertions are evaluated against that episode's private state and recorded dialogue and tool trace (Section~\ref{sec:act:reward}).
\end{document}